\def\year{2022}\relax
\documentclass[letterpaper]{article} % DO NOT CHANGE THIS
\usepackage{aaai22}  % DO NOT CHANGE THIS
\usepackage{times}  % DO NOT CHANGE THIS
\usepackage{helvet}  % DO NOT CHANGE THIS
\usepackage{courier}  % DO NOT CHANGE THIS
\usepackage[hyphens]{url}  % DO NOT CHANGE THIS
\usepackage{graphicx} % DO NOT CHANGE THIS
\usepackage{natbib}  % DO NOT CHANGE THIS AND DO NOT ADD ANY OPTIONS TO IT
\usepackage{caption} % DO NOT CHANGE THIS AND DO NOT ADD ANY OPTIONS TO IT
\DeclareCaptionStyle{ruled}{labelfont=normalfont,labelsep=colon,strut=off} % DO NOT CHANGE THIS
\usepackage{algorithm}
\usepackage{algorithmic}
\usepackage{booktabs} % added by Lasse
\usepackage{pifont} % added by Lasse
\usepackage[dvipsnames]{xcolor} % added by Lasse
\usepackage{array} % added by Lasse
\usepackage{stmaryrd} % added by Lasse
\usepackage{amsmath, amssymb} % added by Lasse
\usepackage{dsfont} % added by Lasse
\usepackage{multirow} % added by Lasse
\usepackage{tikz} % added by Jakob
\usetikzlibrary{bayesnet}
\usetikzlibrary{arrows}
\usetikzlibrary{shapes.multipart}
\usetikzlibrary{positioning}
\usetikzlibrary{calc}
\tikzset{
state/.style={
   rectangle split,
   rectangle split parts=2,
   rectangle split part fill={red!30,blue!20},
   rounded corners,
   draw=black, very thick,
   minimum height=2em,
   text width=3cm,
   inner sep=2pt,
   text centered,
   }
}
\definecolor{customgreen} {RGB}{217	234	212}
\definecolor{customblue}  {RGB}{205	226	242}
\definecolor{customorange}{RGB}{254	228	207}
\definecolor{customred}{RGB}{222 157 155}
\definecolor{sharedcolour}{RGB}{254	228	207}
\usetikzlibrary{shapes}
\usetikzlibrary{fit}
\usetikzlibrary{chains}
\usetikzlibrary{arrows}

\tikzstyle{latent} = [circle,fill=customblue,draw=black,inner sep=1pt,
minimum size=25pt, font=\fontsize{8}{8}\selectfont, node distance=0.75]
\tikzstyle{obs} = [latent,fill=customgreen]
\tikzstyle{const} = [rectangle, inner sep=0pt, node distance=1]
\tikzstyle{factor} = [rectangle, fill=black,minimum size=5pt, inner
sep=0pt, node distance=0.4]
\tikzstyle{det} = [latent, diamond, minimum size=35pt]

\tikzstyle{plate} = [draw, rectangle, rounded corners, fit=#1]
\tikzstyle{wrap} = [inner sep=0pt, fit=#1]
\tikzstyle{gate} = [draw, rectangle, dashed, fit=#1]

\tikzstyle{caption} = [font=\footnotesize, node distance=0] %
\tikzstyle{plate caption} = [caption, node distance=0, inner sep=0pt,
below left=5pt and 0pt of #1.south east] %
\tikzstyle{factor caption} = [caption] %
\tikzstyle{every label} += [caption] %

\tikzset{>={triangle 45}}

\definecolor{xred}{HTML}{C91E12}
\newcommand{\cmark}{{\color{ForestGreen}\ding{51}}}%
\newcommand{\xmark}{{\color{xred}\ding{55}}}%

\newcolumntype{C}[1]{>{\centering\arraybackslash}p{#1}}
\DeclareMathOperator*{\argmax}{arg\,max}

\newcommand{\norm}[1]{\left\lVert#1\right\rVert}

\usepackage{rotating}
\usepackage{makecell}

\usepackage[caption=false,font=normalsize,labelfont=sf,textfont=sf]{subfig}
\usepackage{graphics}

\usepackage{newfloat}
\usepackage{listings}
\floatstyle{ruled}
\newfloat{listing}{tb}{lst}{}
\floatname{listing}{Listing}
\title{A Brief Overview of Unsupervised\\Neural Speech Representation Learning}

\author{
    Lasse Borgholt, \textsuperscript{\rm 1,\rm 3}
    Jakob D. Havtorn, \textsuperscript{\rm 2,\rm 3}
    Joakim Edin, \textsuperscript{\rm 3}
    Lars Maaløe, \textsuperscript{\rm 2,\rm 3}
    Christian Igel \textsuperscript{\rm 1, \rm 4}
}
\affiliations{
    \textsuperscript{\rm 1}Department of Computer Science, University of Copenhagen, Denmark\\
    \textsuperscript{\rm 2}Department of Applied Mathematics and Computer Science, Technical University of Denmark\\
    \textsuperscript{\rm 3}Corti, Copenhagen, Denmark\\
    \textsuperscript{\rm 4}Pioneer Centre for AI, Copenhagen, Denmark\\
    borgholt@di.ku.dk
}
\begin{document}

\maketitle

\begin{abstract}
Unsupervised representation learning for speech processing has matured greatly in the last few years. Work in computer vision and natural language processing has paved the way, but speech data offers unique challenges. As a result, methods from other domains rarely translate directly. We review the development of unsupervised representation learning for speech over the last decade. We identify two primary model categories: self-supervised methods and probabilistic latent variable models. We describe the models and develop a comprehensive taxonomy. Finally, we discuss and compare models from the two categories.
\end{abstract}

\section{Introduction}
Representation learning has shaped modern computer vision \cite{simonyan2014very} and natural language processing \cite{devlin2019bert}, and more recently speech processing has been subject to the same development \cite{baevski2020wav2vec}. Representation learning has been defined as \emph{``learning representations of the data that make it easier to extract useful information when building classifiers or other predictors"} \cite{bengio_representation_2013}.  
Unsupervised representation learning is concerned with learning useful representations without the use of human annotations. Usually, a model is first pre-trained on a task where plenty of data is available. The model is then fine-tuned, or used to extract input representations for a smaller model, targeting a task with limited training data. In computer vision, both supervised \cite{simonyan2014very, szegedy2015going, he2016deep} and unsupervised \cite{pathak2016context, doersch2015unsupervised} representation learning have gained attention with supervised representation learning driven by the availability of large annotated datasets \cite{deng2009imagenet}. For text and speech, pre-training is usually unsupervised as labeled data is difficult to obtain. Although work on text has paved the way, and the two fields share many characteristics, learning representations from speech is a problem faced with a unique set of challenges.

In this paper, we survey work on unsupervised representation learning for speech processing from within the last decade.
From a methodological perspective, we identify two primary model categories, namely models based on self-supervised learning and probabilistic latent variable models. 
We provide a methodological review of the design choices related to each of the model categories and develop a model taxonomy that highlights the different directions of work. Finally, we compare and discuss models from the two categories and their respective evaluation procedures.
%We then provide a comparison of the methods based on their respective evaluation procedures.
%Finally, based on the model taxonomy, we discuss the differences between models from the two categories and identify future challenges and directions for research in speech representation learning. 

\begin{figure}[!t]
    \vspace{4px}
    \begin{flushleft}
        \textsc{\footnotesize \hspace{0.77cm} self-supervised \hspace{1.32cm} latent variable \vspace{0.13cm}}
    \end{flushleft}
    \centering
    \includegraphics[width=0.47\textwidth]{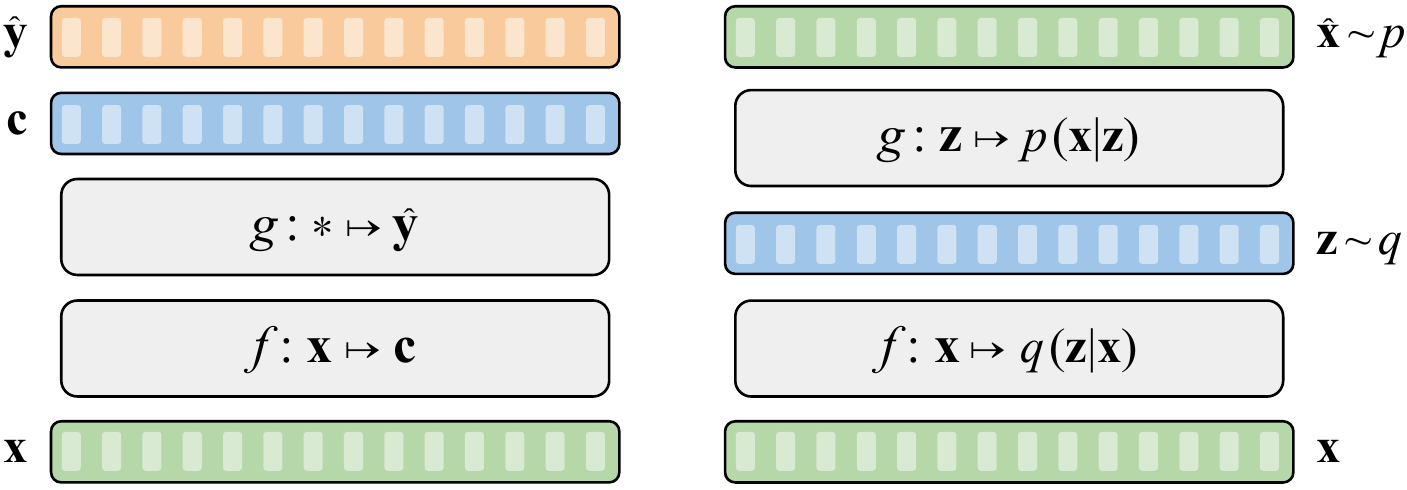}
    \caption{A schematic overview of the two groups of models covered in this survey. \textit{Left}: A model trained with self-supervised learning. We take these models to consist of two functions $f(\cdot)$ and $g(\cdot)$ (sec.~\ref{ssec:notation}). After pre-training, $f(\cdot)$ is fine-tuned or used for extracting features $\mathbf{c}$. $g(\cdot)$ is an auxiliary function used to accommodate the self-supervised pre-training task. \textit{Right}: A probabilistic latent variable model. In contrast to the self-supervised model, the functions $f(\cdot)$ and $g(\cdot)$ learn the parameters of distributions $q$ and $p$. The latent variable $\mathbf{z}$ is commonly used for representation learning.
    }
    \label{fig:ssl_lvm}
\end{figure}

\iffalse
Our contributions are as follows:

\begin{enumerate}
\setlength\itemsep{0mm}
\item We provide a comprehensive review of the development of unsupervised speech representation learning focusing on self-supervised methods and probabilistic latent variable models.
\item Derived from the methodological review, we provide a model taxonomy which we use to discuss differences between models from the two categories.
\item We provide a comparison of the methods based on their respective evaluation procedures and discuss performance, comparability and future research directions.
% \item We identify interesting directions for research within unsupervised representation learning for speech. 
\end{enumerate}
\fi
\section{Unsupervised representation learning}

In the following, we group previous work into \textit{self-supervised models} and \textit{probabilistic latent variable models}, and take a \emph{model} to comprise a neural architecture and a corresponding learning algorithm. A schematic overview is found in figure \ref{fig:ssl_lvm}. These categories are neither exhaustive nor mutually exclusive, but allow us to focus on the characteristics that have shaped different branches of research. 

With emphasis on recent successes in the field, we cover literature from the last 10 years. While a complete description of all relevant models is not within the scope of this work, we sketch important technicalities when they are particularly descriptive of certain models. We first define our high-level notation and conventions to ease discussion.

%Similarly, $\textbf{a}_{<t}$ denotes a vector-sequence containing all elements prior to step $t$ and $\textbf{a}_{t \in \mathbb{A}}$ is a vector sequence containing elements with index contained in the set $\mathbb{A}$.
%(or a scalar sequence $a_{i:j}$)

%Input to a model is either a spectrogram $\mathbf{x}_{1:T}$ or the raw speech signal $x_{1:L}$ where $L > T$. Commonly $L = T(\frac{S}{100})$ where $S$ is the sample rate of the raw speech signal.

% \IEEEpubidadjcol % related to the IEEEpubid command in the top

\subsubsection{Notation} \label{ssec:notation} We use the subscript $i{:}j$ with $i{\leq}j$ to denote a vector sequence $\textbf{a}_{i:j}$ containing elements $\textbf{a}_{i}$ through $\textbf{a}_{j}$. We denote model input as $\mathbf{x}_{1:T}$ which, in practice, might be either a spectrogram or the raw speech signal, but we do not distinguish between the two in notation as it is not essential to understand the models. Also, models commonly downsample the temporal dimension, but again, this is not crucial to understand the models, so we maintain a notation based on a single temporal dimension $t\in\{1,\dots, T\}$.

When discussing self-supervised models, we use $\mathbf{c}_{1:T}$ to denote a contextualized representation. For stochastic latent variable models, we use $\mathbf{z}_{1:T}$ as is customary to the field. 
% The representation at a single time-step, $\mathbf{c}_t$ or $\mathbf{z}_t$, is typically a function of a larger part of the input space than just $\mathbf{x}_t$, in some cases the entire input sequence $\mathbf{x}_{1:T}$.
While some models are frozen and produce representations used as input for downstream tasks (\textbf{\textsc{frz}}, table \ref{tab:model-taxonomy}), others are designed to be fine-tuned (\textbf{\textsc{ftn}}, table \ref{tab:model-taxonomy}). In either case, we use $f(\cdot)$ to denote the model that is used for the downstream task. We use $g(\cdot)$ to denote any auxiliary model components (e.g., for a reconstruction task we might have $g: \mathbf{c}_t \mapsto \hat{\mathbf{x}}_t$). When a model can be naturally subdivided into multiple components, we simply use $f_*(\cdot)$ where $*$ may be any convenient descriptor. Finally, we often use a subscript when defining a loss, $\mathcal{L}_i$, to imply that the total loss is computed as a sum over $i$.

\subsection{Self-supervised models}
\label{sec:ssmodels}

Self-supervised learning is a subset of unsupervised learning \cite{tsai2020self}. Where other unsupervised methods can be seen as a means to an end in itself (e.g., clustering or data generation), self-supervised learning takes the form of a pretext task that only adds value when associated with a downstream task. This makes self-supervised learning tie naturally with semi-supervised learning, but it may also be part of a fully unsupervised setup \cite{baevski2021unsupervised}. Self-supervised learning is often characterized by automatically deriving the target from the input or other unlabeled examples \cite{ouali2020overview}.

\subsubsection{Predictive models} Similar to classic autoregressive language models \cite{mikolov2010recurrent}, contextualized speech representations can be learned by predicting future values of a simple representation \cite{oord2018representation, chung2019unsupervised, schneider2019wav2vec, chung2020generative, jiang2021further} (\textbf{\textsc{prd}}, table \ref{tab:model-taxonomy}). Modeling spectrograms directly, autoregressive predictive coding (APC, \citealp{chung2019unsupervised}) is perhaps the simplest example in this category. The forward pass and loss are computed as
\begin{align}
    \mathbf{c}_{t} &= f(\mathbf{x}_{1:t})  \label{eq:apc_f} \\ 
    \hat{\mathbf{x}}_{t+k} &= g(\mathbf{c}_{t}) \\
    %\mathcal{L}_t(\mathbf{\hat{x}}_{t+k}, \mathbf{x}_{t+k}) &= |\mathbf{\hat{x}}_{t+k} - \mathbf{x}_{t+k}|
    \mathcal{L}_t &= \lVert \hat{\mathbf{x}}_{t+k} - \mathbf{x}_{t+k} \rVert_1\enspace.
\end{align}
\noindent Here, $f(\cdot)$ and $g(\cdot)$ are parameterized by neural networks such that each $\mathbf{c}_t$ is only conditioned on previous inputs $\mathbf{x}_{1:t}$ and $\hat{\mathbf{x}}_{t+k}$ is computed step-wise. \citet{chung2019unsupervised} use a stack of unidirectional LSTMs for $f(\cdot)$ and a linear regression layer for $g(\cdot)$. Tasks that seek to predict or reconstruct the input are very common. In the literature, these are often jointly referred to as reconstruction tasks (\textbf{\textsc{rec}}, table \ref{tab:model-taxonomy}) \cite{liu2021tera, wang2021unispeech}, although this is somewhat misleading in the case of prediction.

Contrary to generative models, such as WaveNet \cite{oord_wavenet:_2016}, the APC model is not restricted to next-step prediction. Instead, it predicts $k > 0$ steps ahead in order to ensure that the model does not learn a trivial solution by exploiting the smoothness of the signal. Depending on the downstream task, we are often interested in learning so-called slow features that will typically span multiple input frames \cite{wiskott2002slow}. Even the smallest linguistic units of speech, phonemes, tend to span $0.1$ seconds on average \cite{garofolo_timit_1993}, whereas spectrogram frames $\mathbf{x}_t$ are typically computed at $0.01$ second intervals. However, sometimes local smoothness is explicitly used to define the task \cite{badino2014auto, jati2017speaker2vec, jati2019neural}.

%In early work, \citet{badino2014auto} propose a step-wise bottleneck model that learns to predict $\mathbf{x}_{t+1}$ from $\mathbf{x}_{t}$. Since phonetic information changes slowly, $\mathbf{x}_{t+1}$ is argued to be a noisy version of $\mathbf{x}_{t}$ encoding almost the same phonetic content. Thus, the model is akin to a denoising autoencoder.
%A similar idea has been proposed by Jati and Georgiou \cite{jati2019neural} to learn speaker embeddings.

% Next-step prediction has also been frequently used for generative models (\textbf{\textsc{gen}}, table \ref{tab:model-taxonomy}). The popular speech synthesis model WaveNet is perhaps the most obvious example \cite{oord_wavenet:_2016}. While this work does not consider representation learning, the model might learn useful features similar to those learned with APC.

\subsubsection{Contrastive models} Speech contains localized noise (e.g., phase shifts) that does not inform slow feature learning. Thus, directly modeling speech might not be the best way to learn contextualized representations. Contrastive predictive coding (CPC,  \citealp{oord2018representation}) targets a local variable $\mathbf{v}_{1:T}$, learned from the model input $\mathbf{x}_{1:T}$, instead of the input itself. The forward pass is
\begin{align}
    \mathbf{v}_{t} &= f_{v}(\mathbf{x}_{t-r:t+r}) \label{eq: cpc local representation} \\
    \mathbf{c}_{t} &= f_{c}(\mathbf{v}_{1:t}) \\
    \hat{\mathbf{v}}_{t,k} &= g_k(\mathbf{c}_{t}) \enspace,
    %g_k(\mathbf{c}_{t}, \mathbf{v}_{t+k}) &= \mathbf{v}_{t+k}^{\intercal} \mathbf{W}_k \mathbf{c}_{t}
    %\mathcal{L}_{t,k} &= - \log \left(\frac{g_k(\mathbf{c}_{t}, \mathbf{v}_{t+k})}{\sum_{n \sim \mathcal{D}} g_k(\mathbf{c}_{t}, \mathbf{v}_{n})} \right)
    % \mathcal{L}_{t,k} &= -\log g_k + \log \sum_{n \sim \mathcal{D}} g_k
\end{align}
\noindent where $f_v(\cdot)$ is a convolutional neural network, such that each $\mathbf{v}_{t}$ only encodes information from a limited receptive field $2r+1$. Again, $f_c(\cdot)$ should be limited to condition each $\mathbf{c}_t$ on previous time-steps $\mathbf{v}_{1:t}$ and $g_k(\cdot)$ is a step-wise transformation. The loss is based on noise constrastive estimation \cite{gutmann2010noise} and is given by
\begin{align}
    \mathcal{L}_{t,k} &= - \log \left(\frac{\exp(\hat{\mathbf{v}}_{t,k}^{\text{\tiny T}}\mathbf{v}_{t+k})}{\sum_{n \sim \mathcal{D}} \exp(\hat{\mathbf{v}}_{t,k}^{\text{\tiny T}}\mathbf{v}_{n})} \right)\enspace .
    % \mathcal{L}_{t,k} &= - \hat{\mathbf{v}}_{t,k}^{\text{\tiny T}}\mathbf{v}_{t+k} + \log \sum_{n \sim \mathcal{D}} \exp(\hat{\mathbf{v}}_{t,k}^{\text{\tiny T}}\mathbf{v}_{n})\enspace .
    % \mathcal{L}_{t,k} &= -\log g_k + \log \sum_{n \sim \mathcal{D}} g_k
    \label{eq: cpc loss}
\end{align}
\noindent Here, $\mathcal{D}$ is a set of indices including the target index $t+k$ and negative samples drawn from a proposal distribution, which is typically taken to be a uniform distribution over the set $\{1,\dots,T\}$. Note that the loss is also indexed by $k$ to show that CPC targets multiple offsets. The APC model is easily extended in a similar way \cite{chung2020improved}.

Crucially, we cannot simply predict $\mathbf{v}_{t+k}$ from $\mathbf{c}_t$ with an $\ell_1$ loss. This would cause $f_v(\cdot)$ to collapse to a trivial solution, such as setting all $\mathbf{v}_t$ equal. With a contrastive loss on the other hand, setting all $\mathbf{v}_t$ equal would cause $\mathcal{L}_{k,t}$ to be constant at a value no better than a random baseline. 

A model closely related to the original CPC model is wav2vec \cite{schneider2019wav2vec}. It uses a different parameterization of the functions $f_v(\cdot)$ and $f_c(\cdot)$, and modifies the loss to consider a binary prediction task, such that we have
\begin{align}
    %g_k(\mathbf{c}_{t}, \mathbf{v}_{t+k}) &= \log(\sigma(\mathbf{v}_{t+k}^{\intercal} (\mathbf{W}_k \mathbf{c}_{t} + \mathbf{b}_k)))\\
    \mathcal{L}_{t,k} &= - \log(\sigma(\hat{\mathbf{v}}_{t,k}^{\text{\tiny T}}\mathbf{v}_{t+k})) - \sum_{n \sim \mathcal{D}} \log(\sigma(-\hat{\mathbf{v}}_{t,k}^{\text{\tiny T}}\mathbf{v}_{n})) \,.
    \label{eq: wav2vec loss}
\end{align}

\noindent This model was among the first to show that learned representations can be used to improve end-to-end speech recognition. As we will see, the wav2vec framework has evolved to shape state-of-the-art representation learning for speech.

\subsubsection{Masking-based models} One downside of predictive tasks is that models are primarily unidirectional. Some work has extended APC and CPC inspired models with separate encoders operating in opposite directions \cite{ling2020deep, kawakami2020learning, borgholt2021scaling}, but these models are still restricted to process left and right context separately. Inspired by the masked language model task used for text-based representation learning \cite{devlin2019bert}, several papers have used masking to overcome this challenge (\textbf{\textsc{msk}}, table \ref{tab:model-taxonomy}). Masking refers to replacing parts of the input with zeros or a learned masking vector. For zero-masking \cite{jiang2019improving, liu2020mockingjay, wang2020unsupervised, chi2021audio, ling2020decoar}, we have
\begin{align}
    \mathbf{c}_{1:T} &= f(\mathbf{x}_{1:T} \circ \mathbf{m}_{1:T})\\
    \mathbf{\hat{x}}_{t} &= g(\mathbf{c}_{t}) \\
    \mathcal{L}_t &= \lVert \mathbf{\hat{x}}_{t} - \mathbf{x}_{t} \rVert_1\enspace,
\end{align}
where the $\circ$ operator denotes the Hadamard product, $f(\cdot)$ is typically a transformer encoder or a bidirectional recurrent neural network, $g(\cdot)$ is a step-wise transformation, and $\mathbf{m}_{1:T}$ is a mask such that $m_{t,i} \in \{0,1\}$. Alternatively, $\mathbf{m}_{1:T}$ is used to select which $\mathbf{x}_t$ are replaced by a learned masking vector. The entries of $\mathbf{m}_{1:T}$ are determined by some stochastic policy. One frequent inspiration is SpecAugment \cite{park2019specaugment}, which was originally proposed for supervised speech recognition and applies frequency and time masking to spectrogram representations.
While temporal masking is most common, frequency masking has also been adopted for representation learning  \cite{wang2020unsupervised}. A simple, yet popular, masking strategy is to draw a proportion of input indices $t_i\sim\{1,\dots,T-M\}$ without replacement, and then mask $\{t_i, \dots, t_i+M\}$ \cite{baevski2020wav2vec, hsu2021hubert, ling2020decoar}.

Combining masking with a contrastive loss, wav2vec 2.0 was the first work to show that a competitive speech recognition model can be learned by fine-tuning a pre-trained model with as little as 10 minutes of labeled data. For this model
\begin{align}
    \mathbf{v}_{t} &= f_v(\mathbf{x}_{t-r:t+r}) \\
    \mathbf{c}_{1:T} &= f_c(\mathbf{v}_{1:T} \circ \mathbf{m}_{1:T}) \\
    \mathbf{q}_t &= g_q(\mathbf{v}_t) \enspace . \label{w2v2 qtz}
\end{align}

\noindent Here, $f_v(\cdot)$ is a convolutional neural network, $f_c(\cdot)$ is a transformer encoder \cite{vaswani2017attention} and $g_q(\cdot)$ is a quantization module used to learn targets from the localized variable $\mathbf{v}_{1:T}$. Computing quantized targets this way requires an extra loss term, which we will present when we discuss quantization in general below. The contrastive loss for wav2vec 2.0 is similar to that of the CPC model,
\begin{align}
    \mathcal{L}_t &= - \log \left(\frac{\exp(S_{\text{c}}(\mathbf{c}_{t}, \mathbf{q}_{t}))}{\sum_{n \sim \mathcal{D}} \exp(S_{\text{c}}(\mathbf{c}_{t}, \mathbf{q}_{n}))} \right) \enspace , \label{w2v2 loss}
\end{align}
\noindent where $S_{\text{c}}(\cdot)$ is the cosine similarity and the negative samples in $\mathcal{D}$ are sampled from other masked time-steps. 

In general, masking is less data efficient than prediction, as only the masked portion of the input is non-trivial to reconstruct. For this reason, the loss might be computed as
\begin{align}
    \mathcal{L}_t &= \lVert (\mathbf{\hat{x}}_{t} - \mathbf{x}_{t}) \circ (\mathbf{1} - \mathbf{m}_t) \rVert_1 \enspace .
\end{align}

\begin{figure}[!t]
    \centering
    \setlength\tabcolsep{1.5pt}
    \begin{tabular}{>{\centering\arraybackslash} m{4mm}  >{\centering\arraybackslash} m{0.19\textwidth}|>{\centering\arraybackslash} m{0.19\textwidth}}
          & {\footnotesize \textsc{predictive}} & {\footnotesize \textsc{masking-based}} \\
        %\midrule
        \rotatebox{90}{{\footnotesize \textsc{reconstruct}}} & \includegraphics[width=0.19\textwidth]{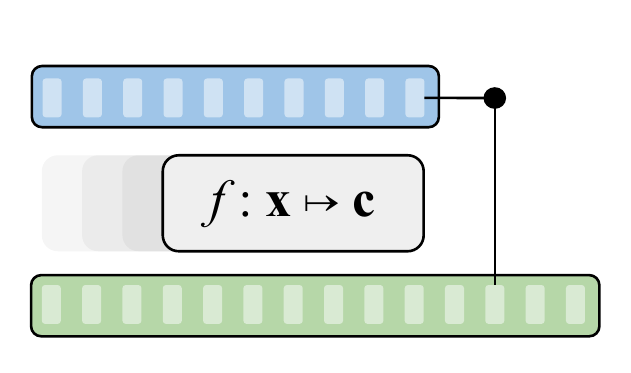} & \includegraphics[width=0.19\textwidth]{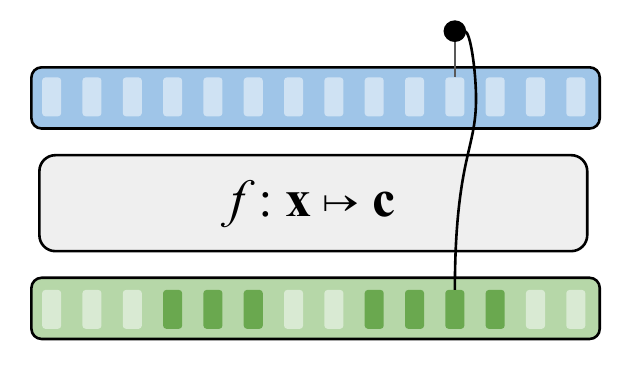}  \\
        \midrule
        \rotatebox{90}{{\footnotesize \textsc{contrastive}}} & \includegraphics[width=0.19\textwidth]{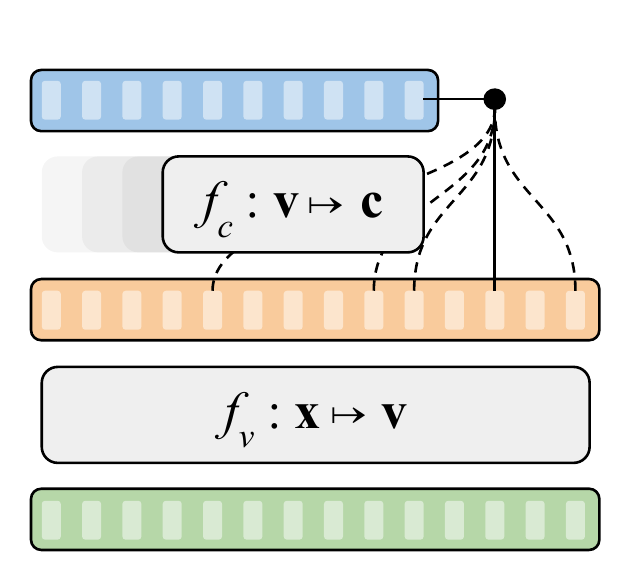} & \includegraphics[width=0.19\textwidth]{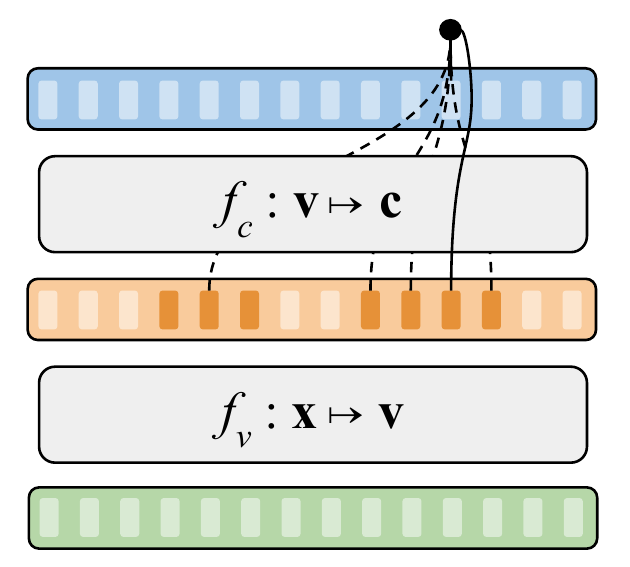}
    \end{tabular}
    \caption{
    Schematic of self-supervised methods. Each subfigure illustrates the loss computation for a single time-step.
    The temporal subscript has been left out for simplicity. 
    %Examples of models in each category include APC (\textbf{\textsc{prd}} + \textbf{\textsc{rec}}), Mockingjay (\textbf{\textsc{msk}} + \textbf{\textsc{rec}}), CPC (\textbf{\textsc{prd}} + \textbf{\textsc{con}}) and wav2vec 2.0 (\textbf{\textsc{msk}} + \textbf{\textsc{con}}).
    }
    \label{fig:ssl_grid}
\end{figure}

\noindent Non-autoregressive predictive coding (NPC, \citealp{liu2020non}) tries to resolve this by using a convolutional neural network where the kernel is masked instead of the input.
%That is, the model is prevented from using $\mathbf{x}_{t}$ for its corresponding reconstruction $\mathbf{\hat{x}}_{t}$.
This allows for complete data utilization, but limits the amount of context encoded in the learned representation.
Figure \ref{fig:ssl_grid} summarizes the models discussed so far.

\begin{table*}[t]
\caption{Selected models classified according to the binary attributes identified throughout the text. The models are sorted according to first publication date on arXiv which might differ from the citation year. \textbf{\textsc{msk}}: masking, \textbf{\textsc{prd}}: prediction, \textbf{\textsc{con}}: contrastive, \textbf{\textsc{rec}}: reconstruction, \textbf{\textsc{qtz}}: quantization, \textbf{\textsc{gen}}: generative, \textbf{\textsc{frz}}: frozen, \textbf{\textsc{ftn}}: fine-tuned, \textbf{\textsc{loc}}: local, \textbf{\textsc{glo}}: global.\vspace{-1px}
}

%\textbf{\textsc{msk}} means that the model uses masking, \textbf{\textsc{prd}} that it uses prediction, \textbf{\textsc{con}} that it uses a contrastive loss, \textbf{\textsc{rec}} that it uses a reconstruction loss, \textbf{\textsc{qtz}} that it uses quantization, \textbf{\textsc{gen}} that it is generative, \textbf{\textsc{frz}} that frozen representations are extracted for downstream tasks, \textbf{\textsc{ftn}} that the model is fine-tuned for downstream tasks, \textbf{\textsc{loc}} that the model learns local representations and \textbf{\textsc{glo}} that it learns a single global representation.
\label{tab:model-taxonomy}
\begin{center}
\renewcommand{\arraystretch}{1.1}
\resizebox{0.925\textwidth}{!}{%
\begin{tabular}{ l l | 
c c c c c c | c c c | c c } 
\toprule

%& & & & & & \multicolumn{2}{c}{\textbf{\textsc{architecture}}} \\

\multicolumn{2}{c}{} & 
\multicolumn{6}{c}{\footnotesize\textsc{Model and task design}} & 
\multicolumn{3}{c}{\footnotesize\textsc{Resolution}} &
\multicolumn{2}{c}{\footnotesize\makebox[0pt][c]{\textsc{Usage}}} \\
\textbf{\textsc{model}} &
\textbf{\textsc{pub. date}} &
\textbf{\textsc{msk}} & 
\textbf{\textsc{prd}} & 
\textbf{\textsc{con}} &  
\textbf{\textsc{rec}} &  
\textbf{\textsc{qtz}} & 
\textbf{\textsc{gen}} & 
\textbf{\textsc{loc}} & 
\textbf{\textsc{glb}} & 
\textbf{\textsc{var}} & 
\textbf{\textsc{frz}} & 
\textbf{\textsc{ftn}} \\

\midrule
\multicolumn{12}{c}{\textsc{Self-supervised models}} \\
\midrule

%                                                           MSK      PRD      CON      REC      QTZ      GEN      LOC      GLB     VAR      FRZ      FTN
\textbf{Audio Word2vec} \footnotesize\cite{chung2016audio}      & 2016 Mar. & \cmark & \xmark & \xmark & \cmark & \xmark & \xmark & \xmark & \cmark & \xmark & \cmark & \xmark \\
\textbf{Speech2Vec} \footnotesize\cite{chung2018speech2vec}     & 2018 Mar. & \xmark & \cmark & \xmark & \cmark & \xmark & \xmark & \xmark & \cmark & \xmark & \cmark & \xmark \\
\textbf{Unspeech} \footnotesize\cite{milde2018unspeech}         & 2018 Apr. & \xmark & \cmark & \cmark & \xmark & \xmark & \xmark & \xmark & \cmark & \xmark & \cmark & \xmark \\

\textbf{CPC} (van den Oord et al. 2018)         & 2018 Jul. & \xmark & \cmark & \cmark & \xmark & \xmark & \xmark & \cmark & \xmark & \xmark & \cmark & \xmark \\

%PASE \footnotesize\cite{Pascual2019}                   & 2019 Apr. & \xmark & \xmark & \cmark & \cmark & \xmark & \xmark & \cmark & \xmark & \xmark & \cmark & \cmark \\ % 6/4
\textbf{APC} \footnotesize\cite{chung2019unsupervised}          & 2019 Oct. & \xmark & \cmark & \xmark & \cmark & \xmark & \xmark & \cmark & \xmark & \xmark & \cmark & \xmark \\ % 5/4
\textbf{wav2vec} \footnotesize\cite{schneider2019wav2vec}       & 2019 Apr. & \xmark & \cmark & \cmark & \xmark & \xmark & \xmark & \cmark & \xmark & \xmark & \cmark & \xmark \\ % 11/4

%VQ-APC \footnotesize\cite{chung2020vector}             & 2020 May  & \xmark & \cmark & \xmark & \cmark & \cmark & \xmark & \cmark & \xmark & \xmark & \cmark & \xmark \\ 

\textbf{Mockingjay}  \footnotesize\cite{liu2020mockingjay}      & 2019 Oct. & \cmark & \xmark & \xmark & \cmark & \xmark & \xmark & \cmark & \xmark & \xmark & \cmark & \cmark \\ % 25/10-19
%DeCoAR \footnotesize\cite{ling2020deep}                & 2019 Dec. & \xmark & \cmark & \xmark & \cmark & \xmark & \xmark & \cmark & \xmark & \xmark & \cmark & \xmark \\ % 3/12-19
%PASE+ \footnotesize\cite{Ravanelli2020}                & 2020 Jan. & \cmark & \xmark & \cmark & \cmark & \xmark & \xmark & \cmark & \xmark & \xmark & \cmark & \cmark \\ % 25/1
\textbf{wav2vec 2.0} \footnotesize\cite{baevski2020wav2vec}     & 2020 Jun. & \cmark & \xmark & \cmark & \xmark & \cmark & \xmark & \cmark & \xmark & \xmark & \xmark & \cmark \\ % 20/6
%TERA \footnotesize\cite{liu2021tera}                   & 2021 Jul. & \cmark & \xmark & \xmark & \cmark & \xmark & \xmark & \cmark & \xmark & \xmark & \cmark & \cmark \\ % 12/7
\textbf{NPC} \footnotesize\cite{liu2020non}                     & 2020 Nov. & \cmark & \xmark & \xmark & \cmark & \cmark & \xmark & \cmark & \xmark & \xmark & \cmark & \xmark \\ % 1/11
\textbf{DeCoAR 2.0} \footnotesize\cite{ling2020decoar}          & 2020 Dec. & \cmark & \xmark & \xmark & \cmark & \cmark & \xmark & \cmark & \xmark & \xmark & \cmark & \xmark \\ % 11/12
\textbf{SCPC} \footnotesize\cite{bhati2021segmental}            & 2021 Jun. & \xmark & \cmark & \cmark & \xmark & \xmark & \xmark & \cmark & \xmark & \cmark & \cmark & \xmark \\ % 3/6
\textbf{HuBERT} \footnotesize\cite{hsu2021hubert}               & 2021 Jun. & \cmark & \xmark & \xmark & \xmark & \cmark & \xmark & \cmark & \xmark & \xmark & \xmark & \cmark \\ % 14/6

\midrule
\multicolumn{12}{c}{\textsc{Probabilistic latent variable models}} \\
\midrule
%                                                           MSK      PRD      CON      REC      QTZ      GEN      LOC      GLO     VAR      FRZ      FTN
% Conv. DBN \footnotesize\cite{lee_unsupervised_2009}    & 2009 Jan. & \xmark & \xmark & \xmark & \cmark & \xmark & \cmark & \cmark & \xmark & \cmark & \xmark & \xmark \\
\textbf{VRNN} \footnotesize\cite{chung_recurrent_2015}          & 2015 Jun. & \xmark & \xmark & \xmark & \cmark & \xmark & \cmark & \cmark & \xmark & \xmark & \cmark & \xmark \\
\textbf{SRNN} \footnotesize\cite{fraccaro_sequential_2016}      & 2016 May  & \xmark & \xmark & \xmark & \cmark & \xmark & \cmark & \cmark & \xmark & \xmark & \cmark & \xmark \\
\textbf{HMM-VAE} \footnotesize\cite{ebbers_hidden_2017}         & 2017 Mar. & \xmark & \xmark & \xmark & \cmark & \xmark & \cmark & \cmark & \xmark & \xmark & \cmark & \xmark \\
\textbf{ConvVAE} \footnotesize\cite{hsu_learning_2017}          & 2017 Apr. & \xmark & \xmark & \xmark & \cmark & \xmark & \cmark & \xmark & \cmark & \xmark & \cmark & \xmark \\
\textbf{FHVAE} \footnotesize\cite{hsu_unsupervised_2017}        & 2017 Sep. & \xmark & \xmark & \xmark & \cmark & \xmark & \cmark & \cmark & \cmark & \xmark & \cmark & \xmark \\
\textbf{VQ-VAE} (van den Oord et al. 2018)            & 2017 Nov. & \xmark & \xmark & \xmark & \cmark & \cmark & \cmark & \cmark & \xmark & \xmark & \cmark & \xmark \\
\textbf{BHMM-VAE} \footnotesize\cite{glarner_full_2018}         & 2018 Mar. & \xmark & \xmark & \xmark & \cmark & \xmark & \cmark & \cmark & \xmark & \xmark & \cmark & \xmark \\
\textbf{STCN} \footnotesize\cite{aksan_stcn_2019}               & 2019 Feb. & \xmark & \xmark & \xmark & \cmark & \xmark & \cmark & \cmark & \xmark & \xmark & \cmark & \xmark \\
\textbf{FDMM} \footnotesize\cite{khurana_factorial_2019}        & 2019 Oct. & \xmark & \xmark & \xmark & \cmark & \xmark & \cmark & \cmark & \cmark & \xmark & \cmark & \xmark \\
\textbf{ConvDMM} \footnotesize\cite{khurana_convolutional_2020} & 2020 Jun. & \xmark & \xmark & \xmark & \cmark & \xmark & \cmark & \cmark & \xmark & \xmark & \cmark & \xmark \\
\bottomrule
\end{tabular}
}
\end{center}
\end{table*}

\subsubsection{Quantization} 
Several models enforce a discrete latent space by quantizing the vector representation (\textbf{\textsc{qtz}}, table \ref{tab:model-taxonomy}). The two most popular approaches are the Gumbel-softmax \cite{jang2016categorical, maddison2016concrete} and the quantization used in the VQ-VAE \cite{oord_neural_2018}.

%Several models enforce a discrete latent space by quantizing the vector representation (\textbf{\textsc{qtz}}, table \ref{tab:model-taxonomy}). A popular approach for online vector quantization is to use the Gumbel softmax \cite{jang2016categorical, maddison2016concrete}. This approach corresponds to approximate differentiable sampling from a categorical distribution. Another common approach employs a codebook of learnable representations. This approach was popularized with the VQ-VAE \cite{oord_neural_2018}, and thus, we refer to it as the VQ-VAE approach. Both approaches are non-differentiable and obtain gradients using the straight-through estimator \cite{bengio2013estimating}.

\textit{Gumbel-softmax approach:} 
Say we want to quantize a vector $\mathbf{v}$ such that it takes one of $K$ possible values. We first map $\mathbf{v}$ to $\mathbf{l} \in \mathbb{R}^K$ and then map $\mathbf{l}$ to a probability vector $\mathbf{p} \in \mathbb{R}^K$ via the Gumbel softmax given by
\begin{align}
    p_i &= \frac{\exp(l_i + n_i) / \tau}{\sum_k^K \exp(l_k + n_k) / \tau} 
\end{align}
% for $i=1,\dots,K$. Here $\tau$ is a temperature parameter and $\mathbf{n}\in\mathbb R^K$ is a random vector with $n_i = -\log(-\log(u_i))$ for $u_i \sim \mathcal{U}(0, 1)$. 
% As  $\tau \rightarrow 0$, $\mathbf{p}$ approaches a one-hot vector. The Gumbel noise $\mathbf{n}$ is a practical way to sample from the untempered categorical distribution (i.e., $\tau = 1$). 
% A function $\varphi(\mathbf{p})$ is then used to map $\mathbf{p}$ to a discrete sample which can then be used to obtain the quantized vector (e.g., $\mathbf{q}_t$ in eq.~\ref{w2v2 qtz}) via a codebook lookup. As this function is non-differentiable, we have to rely on the straight-through estimator \cite{bengio2013estimating} which assumes that the Jacobian ${\partial \, \varphi}/{\partial \, \mathbf{p}}$ equals the identity matrix. 
for $i=1,\dots,K$. Here $\tau$ is a temperature parameter and $\mathbf{n}\in\mathbb R^K$ is a random vector with $n_i = -\log(-\log(u_i))$ for $u_i \sim \mathcal{U}(0, 1)$.
As  $\tau \rightarrow 0$, $\mathbf{p}$ approaches a one-hot vector. The Gumbel noise $\mathbf{n}$ is a practical way to sample from the untempered categorical distribution (i.e., $\tau = 1$).
$\mathbf{p}$ is mapped to a one-hot vector using a function $\varphi(\cdot)$, such that $\varphi(\mathbf{p})_i = 1$ if $i = \argmax_j p_j$ and $0$ otherwise. 
As this function is non-differentiable, we must rely on the straight-through gradient estimator \cite{bengio2013estimating} which assumes that the Jacobian ${\partial \, \varphi}/{\partial \, \mathbf{p}}$ equals the identity matrix. 
The one-hot vector can then be used for a codebook lookup to obtain the final quantized vector (e.g., $\mathbf{q}_t$ in eq.~\ref{w2v2 qtz}).

%We choose $\varphi(\cdot)$ to return a one-hot vector with $\varphi(\mathbf{p})_i = 1$ if $i = \argmax_j p_j$ and $0$ otherwise. The one-hot vector can then be used for a codebook lookup. For non-differentiable $\varphi(\cdot)$, the gradient is usually taken to be that of the input to $\varphi(\cdot)$. That is, we use the straight-through estimator and assume during training that the Jacobian ${\partial \, \varphi}/{\partial \, \mathbf{p}}$ equals the identity matrix. 
%\begin{align}
%    %\mathbf{v} &= \mathbf{p} + \textsc{sg}(\varphi(\mathbf{p}) - %\mathbf{p}) \, .
%    \frac{\partial \, \varphi}{\partial \, \mathbf{p}} = \mathbf{I}
%\end{align}
%where $\textsc{sg}(\cdot)$ is the \textit{stop gradient} function; an identity function that blocks the gradient from the input. .

The wav2vec 2.0 quantization module (eq.~\ref{w2v2 qtz}) uses the Gumbel softmax. To ensure utilization of codebook vectors, a diversity loss is added to the task specific loss (eq. \ref{w2v2 loss})
\begin{align}
    % \mathcal{L} =  - H(\widetilde{\mathbf{p}}) \frac{1}{K}\enspace,
    \mathcal{L} =  - H(\widetilde{\mathbf{p}}) / K \enspace,
\end{align}
\noindent where $H(\cdot)$ is the entropy and $\widetilde{\mathbf{p}}$ is the untempered version of $\mathbf{p}$ without Gumbel noise.

\textit{VQ-VAE approach:} Instead of directly parameterizing a probability distribution, as in the Gumbel softmax, a vector $\mathbf{v}$ can be quantized by replacing it with the closest codebook vector $\mathbf{e}_k$. Specifically, given a learned codebook $\mathbf{e}\in\mathbb{R}^{K\times D}$, where $K$ is the codebook size and $D$ is the dimensionality of each codebook vector $\mathbf{e}_k$, the quantized representation $\mathbf{q}$ of $\mathbf{v}$ is obtained as,
\begin{align}
    \mathbf{q} = \mathbf{e}_k \ , \enspace \text{where} \enspace k=\arg\min_j \norm{\mathbf{v} - \mathbf{e}_j}_2 \enspace .
\end{align}
As $\arg\min$ is non-differentiable, the straight-through estimator is used as for the Gumbel-softmax. 
% The straight-through estimator defines the gradient of this operation by setting the gradient of $\mathbf{v}$ wrt. to the loss equal to that of $\mathbf{q}$. 
%
% Codebook learning is facilitated by a two-term auxiliary loss similar to classical vector quantization dictionary learning \cite{burton_generalization_1983, soong_vector_1985}. Gradients for the codebook vectors are given solely by a vector quantization term, which moves codebook vectors $\mathbf{e}_k$ closer to the non-quantized vectors $\mathbf{v}$. A so-called commitment term is added to ensure that non-quantized vectors do not grow unboundedly by enforcing the encoder to keep them close to a codebook vector.
%
Codebook learning is facilitated by a two-term auxiliary loss similar to classical vector quantization dictionary learning \cite{burton_generalization_1983, soong_vector_1985}. Gradients for the codebook vectors are given solely by a vector quantization term. A so-called commitment term ensures that non-quantized vectors do not grow unboundedly.
\begin{align}
    \mathcal{L} = \underset{\text{vq}}{\underbrace{\norm{\text{sg}\left[\mathbf{v}\right] - \mathbf{e}}_2^2}} + \underset{\text{commitment}}{\underbrace{\beta\norm{\mathbf{v} - \text{sg}\left[\mathbf{e}\right]}_2^2}} \enspace , \label{eq: vector quantization losses}
\end{align}
where $\text{sg}[\mathbf{x}] = \mathbf{x}$ is the stop-gradient operator with the property $\frac{d}{dx_i}\text{sg}[\mathbf{x}] \equiv 0$ for all $i$ and $\beta$ is a hyperparameter. Although vector quantization was introduced by the VQ-VAE which is, in some ways, a latent variable model, it has been applied to self-supervised methods \cite{niekerk_vector_2020, baevski2019vq}.

%but with variations between individual approaches. The VQ-CPC model \cite{niekerk_vector_2020} quantizes $\mathbf{v}$ in the original CPC (eq.~\ref{eq: cpc local representation}) and augments its loss (eq.~\ref{eq: cpc loss}) by addition of the commitment term in eq.~\ref{eq: vector quantization losses}. Instead of learning the codebook via the VQ term, it is updated as a moving average of $\mathbf{v}$ as initially suggested by \citet{oord_neural_2018}. 
%The vq-wav2vec \cite{baevski2019vq} is defined similarly but adds the full auxiliary loss to the wav2vec loss (eq.~\ref{eq: wav2vec loss}) in place of moving averages.

\textit{Motivation:} Similar to how quantization approaches differ between works, so do the motivations provided for employing them. The vq-wav2vec \cite{baevski2019vq, baevski2019effectiveness} %quantizes $\mathbf{v}_{1:T}$ before feeding them to the context network $f_c(\cdot)$. Here, the motivation is 
learn quantized representations in order to apply natural language processing models, like BERT \cite{devlin2019bert}, afterwards. Other works use quantization for speech segmentation \cite{kamper2020towards, chorowski2019unsupervised} or as a bottleneck in order to ``\textit{limit model capacity}" \cite{chung2020vector, ling2020decoar}.
%Quantized representations have also been used for speech segmentation \cite{kamper2020towards, chorowski2019unsupervised}. 
Finally, \citet{chung2020vector} explore quantization between different layers in the APC model, but find that continuous representations consistently perform better than their quantized counterparts on a downstream phoneme classification task

%They find that applying quantization to the output $\mathbf{c}_{1:T}$ yields the best results on a downstream phoneme classification task. However, the continuous representations consistently perform better than their quantized counterparts.

Given our previous discussion of how it might not be beneficial to model localized noise, quantization in wav2vec 2.0 seems well motivated, as it enforces the target representation $\mathbf{q}_{1:T}$ to discard such noise. Taking this idea further, the HuBERT model \cite{hsu2021hubert} uses offline quantization to learn categorical targets. Initially, spectrogram features are used to learn frame-wise labels with $k$-means clustering. A model similar to wav2vec 2.0, but without online quantization, is then trained to infer labels for masked time-steps. Since quantization is offline, this model does not need to rely on a contrastive loss, but can infer the target class directly. The offline quantization also ensures more stable training, as targets do not change abruptly.
\nopagebreak

\subsubsection{Global representations}
The models covered so far learn representations that maintain a temporal resolution proportional to the input resolution. We say that they learn local representations (\textbf{\textsc{loc}}, table \ref{tab:model-taxonomy}). Now, we cover models that learn global representations (\textbf{\textsc{glb}}, table \ref{tab:model-taxonomy}).
%Apart from this attribute, we find that models that learn global representations rely on many of the same techniques described in the previous sections. %We first give a broad overview, before discussing how they relate to the self-supervised models.

Early work on global speech representation learning takes inspiration from the autoencoder framework \cite{kramer1991nonlinear}. \citet{chung2016audio} propose a simple sequence-to-sequence autoencoder for learning acoustic word embeddings:
%
% i/x-vectors here?
\begin{align}
    \mathbf{c} &= f(\mathbf{x}_{1:T}) \\
    \hat{\mathbf{x}}_{1:T} &= g(\mathbf{c}) \enspace , \label{eq:dec-aw2v}
\end{align}

\noindent where $f(\cdot)$ and $g(\cdot)$ are a recurrent neural networks, such that $\mathbf{c}$ is taken to be the hidden state at the last time-step $T$ of $f(\cdot)$ and used as initial hidden state of $g(\cdot)$. 
%where $f(\cdot)$ is a recurrent neural network encoder, such that $\mathbf{c}$ is taken to be the hidden state at the last time-step $T$. The decoder, $g(\cdot)$, is also a recurrent neural network where $\mathbf{c}$ is used to initialize the hidden state.
The authors also propose a denoising autoencoder with masked inputs $f(\mathbf{x}_{1:T} \circ \mathbf{m}_{1:T})$. Similar RNN-based autoencoders have also been explored \cite{kamper2019truly, holzenberger2018learning}.
%
%\begin{align}
%    \mathbf{c} &= f(\mathbf{x}_{1:T} \circ \mathbf{m}_{1:T})
%\end{align}
%

Prior to this work, \citet{kamper2015unsupervised} and \citet{renshaw2015comparison} introduced the \textit{correspondence autoencoder}. This method uses dynamic time warping to align input-target segment pairs extracted with unsupervised term discovery. In more recent work, the need for alignment has been alleviated by adopting the sequence-to-sequence framework \cite{kamper2019truly, jacobs2021acoustic}.

%This method relies on unsupervised term discovery with randomized algorithms \cite{jansen2011efficient} to extract corresponding speech segment pairs for training. Thus, one segment is used as input and the other as reconstruction target. Because the paired segments do not necessarily have the same length, they need to be aligned with dynamic time warping.

% In more recent work, this need has been alleviated by adopting the sequence-to-sequence framework \cite{kamper2019truly, jacobs2021acoustic}.
%Unlike the reconstruction tasks reviewed in the previous sections, models presented here all rely on an $\ell_2$ loss, $\mathcal{L} = \lVert \hat{\mathbf{x}}_{1:T} - \mathbf{x}_{1:T} \rVert_2\enspace$.
%
% \begin{align}
%     \mathcal{L} &= \lVert \hat{\mathbf{x}}_{1:T} - \mathbf{x}_{1:T} \rVert_2\enspace.
% \end{align}

Inspired by the work on semantic word embeddings for text \cite{mikolov2013distributed}, the sequence-to-sequence framework has also been used to implement speech-based versions of the skipgram and continuous bag-of-words models \cite{chung2017learning, chung2018speech2vec}. Given a segment corresponding to a single word $\mathbf{x}_{(n)} = \mathbf{x}_{t_{n}:t_{n+1}}$, the skipgram model is trained to predict neighboring words $\mathbf{x}_{(n + k)}$ where $k\ne0$. That is, instead of a single decoder, as in eq. \ref{eq:dec-aw2v}, the skipgram model employs multiple decoders
\begin{align}
    \hat{\mathbf{x}}_{(n + k)} = g_k(\mathbf{c})\enspace.
\end{align}
Conversely, the continuous bag-of-words model is trained to predict the target word from the neighboring words, so here multiple encoders sum over several offsets $\mathcal{K}$ to obtain $\mathbf{c}$:
\begin{align}
    \mathbf{c} &= \sum_{k\in\mathcal{K}} f_k(\mathbf{x}_{(n + k)}) %\enspace .
    % \mathbf{c} &= \sum_{k\in\mathcal K_n} \mathbf{c}_k
\end{align}
%for different values of $k$.
The sequence-to-sequence models described above rely on speech segments corresponding to words. The segments are obtained by supervised forced alignment, %Thus, since forced alignment requires labeled data, the models might be seen as weakly supervised
but similar models have been explored without this requirement \cite{jati2017speaker2vec, tagliasacchi2020pre}.

Contrastive learning has also been explored for global speech representation learning \citep{milde2018unspeech, jati2019neural, jansen2018unsupervised}. And prior to the widespread adoption of neural networks, \citet{levin2013fixed} explore principal component analysis and Laplacian eigenmaps for learning fixed-sized acoustic embeddings.

\subsubsection{Other work} Some models learn local representations with a variable temporal resolution that is not proportional to the input resolution (\textbf{\textsc{var}}, table \ref{tab:model-taxonomy}). In practice, this is often achieved implicitly, by learning segment boundaries or by taking repeated quantized values to belong to the same segment \cite{kamper2020towards, chorowski2019unsupervised, michel2017blind, kreuk2020self, wang2017gate, dieleman_variable-rate_2021}. An exception is the recently proposed \emph{segmental contrastive predictive coding} (SCPC, \citealp{bhati2021segmental, bhati2021unsupervised}). With this approach, the model explicitly learns segment boundaries, which are used to downsample the representations during training. The same segmentation strategy has subsequently been applied in other models \cite{cuervo2021contrastive}.

Most of the work presented so far fits neatly into the taxonomy presented in table \ref{tab:model-taxonomy}. One exception is the problem-agnostic speech encoder (PASE, \citealp{Pascual2019, Ravanelli2020}) that combines multiple pre-training tasks. Furthermore, many of the presented models have been successfully applied to other use cases. For instance, wav2vec 2.0 and related models have been applied to learn cross-lingual and multi-lingual representations \cite{riviere2020unsupervised, conneau2020unsupervised, khurana2021magic} and proven well-suited for concurrently learning with labeled data \cite{talnikar2021joint, wang2021unispeech}. %We will discuss the downstream application of self-supervised representations further in section \ref{sec:eval}.

\subsection{Probabilistic latent variable models}
\label{sec:plvms}
Another prominent class of models are probabilistic latent variable models (LVMs). 
Before surveying their application to speech, we briefly review LVMs and their usual specification when applied for representation learning in general. We disregard any specific temporal notation without loss of generality. 
We then introduce the variational autoencoder framework (VAE, \citealp{kingma_auto-encoding_2014}). We focus on different dependency structures between data and learned representations, in contrast to the more practical view on self-supervised models taken above.

\subsubsection{LVMs and inference}
Fundamental to LVMs is the assumption that the data is produced by a generative process that involves unobserved stochastic latent variables $\textbf{z}$. 
An LVM aims to model this generative process to enable generation of new data $\mathbf{x}$ (\textbf{\textsc{gen}}, table \ref{tab:model-taxonomy}) and inference of the latent variable associated with a given observed variable $\textbf{x}$. 
For representation learning, the inference of latent variables is of primary interest.
An LVM is defined by the observation model $p(\mathbf{x}|\mathbf{z})$, which defines the relationship between the observed and latent variables, and the prior $p(\mathbf{z})$, which defines the relationship among the latent variables \cite{bartholomew_latent_2011}. 
An LVM models the generative process via the joint observation and prior model $p(\mathbf{x}, \mathbf{z})$ often referred to as the generative model. 
The likelihood of an LVM given an example $\textbf{x}$ can be written as
\begin{equation}
    \log p(\mathbf{x}) = \log \int p(\mathbf{x}|\mathbf{z}) p(\mathbf{z}) \,\text{d}\mathbf{z} \enspace.
    \label{eq: lvm log-likelihood}
\end{equation}
The latent variable can be inferred with e.g. Markov Chain Monte Carlo (MCMC) methods \cite{mohamed_monte_2019} or variational inference \cite{jordan_introduction_1999}.

For representation learning, LVMs are commonly defined using the VAE framework \cite{kingma_auto-encoding_2014, rezende_stochastic_2014}) which is also the focus of our exposition. 
In the VAE framework, the observation model $p(\mathbf{x}|\mathbf{z})$ is parameterized using a deep neural network. This choice allows modeling complex and high-dimensional data but also makes the integral in eq.~\ref{eq: lvm log-likelihood} analytically intractable. MCMC methods can be used to estimate it and the true model posterior $p(\mathbf{z}|\mathbf{x})$, but these methods are usually computationally expensive in this setting \cite{mohamed_monte_2019}. 
To counter this and make gradient-based maximum likelihood training feasible, the VAE instead employs variational inference \cite{jordan_introduction_1999}. It approximates the intractable true model posterior by introducing a variational posterior distribution $q(\mathbf{z}|\mathbf{x})$, also parameterized by a deep neural network. From eq.~\ref{eq: lvm log-likelihood}, via Jensen's inequality, this gives rise to a variational lower bound on the likelihood, also known as the evidence lower bound (ELBO).
\begin{equation}
    \log p(\mathbf{x}) \geq \int q(\mathbf{z}|\mathbf{x}) \log \frac{p(\mathbf{x}| \mathbf{z})p(\mathbf{z})}{q(\mathbf{z}|\mathbf{x})} \,\text{d}\mathbf{z} \equiv \mathcal{L}_{\text{ELBO}} \enspace . \label{eq: lvm likelihood bound (elbo)}
\end{equation}
The bound can be efficiently evaluated and optimized with Monte Carlo (MC) estimation by sampling from $q(\mathbf{z}|\mathbf{x})$. Low-variance gradient estimates are usually obtained via reparameterization of $q(\mathbf{z}|\mathbf{x})$ \cite{kingma_auto-encoding_2014} although alternatives exist (e.g., inverse CDF sampling) \cite{mohamed_monte_2019}. 
The ELBO can also be written as
\begin{equation}
    \mathcal{L}_{\text{ELBO}} = \mathbb{E}_{q(\mathbf{z}|\mathbf{x})}\left[ \log p(\mathbf{x}|\mathbf{z}) \right] - D_\text{KL}\left( q(\mathbf{z}|\mathbf{x}) || p(\mathbf{z}) \right) \enspace  \label{eq: lvm likelihood bound recon/kl form (elbo)},
\end{equation}
where $\mathbb{E}\left[ \log p(\mathbf{x}|\mathbf{z}) \right]$ can be seen as a reconstruction loss and $D_\text{KL}\left( q(\mathbf{z}|\mathbf{x})||p(\mathbf{z}) \right)$ is the Kullback-Leibler (KL) divergence between the variational posterior distribution and the prior.

In brief, LVMs of the VAE-type consist of a approximate posterior, $q(\mathbf{z}|\mathbf{x})$, an observation model, $p(\mathbf{x}|\mathbf{z})$, and a prior, $p(\mathbf{z})$. 
% The joint observation model and prior form the generative model which can be efficiently sampled with ancestral sampling which entails first sampling a latent variable $\tilde{\mathbf{z}}\sim p(\mathbf{z})$ followed by sampling the observed variable conditioned on that latent variable, $\tilde{\mathbf{x}}\sim p(\mathbf{x}|\tilde{\mathbf{z}})$. 
With reference to probabilistic coding theory, the approximate posterior is often referred to as the encoder and the observation model as the decoder \cite{kingma_auto-encoding_2014, rezende_stochastic_2014}. From a theoretical perspective, the encoder exists solely as the result of choosing to use variational inference to train the decoder rather than e.g. MCMC. As such, it is also referred to as the inference model. However, from a representation learning perspective, the encoder is essential as it can be used to efficiently obtain the representation $\mathbf{z}$ commonly used for downstream tasks. 
It is still possible to evaluate and sample the true posterior distribution $p(\mathbf{z}|\mathbf{x})$ by applying MCMC methods such as Hamiltonian Monte Carlo on the decoder, but for computational reasons this is rarely done in practice. 

\begin{table}[t!]
\caption{
A comprehensive overview of observation, prior and inference models for VAE type latent variable models with a single latent variable. 
The observation, prior and inference models may all belong to one or more of the categories listed under them as detailed in section \ref{sec:plvms}. 
The types listed here serve as primitives from which more complex structures can be constructed including models with hierarchies of multiple latent variables. 
% We indicate autoregressiveness (\textbf{\textsc{arx}} and \textbf{\textsc{arz}}) using a ``catch-all" notation $*$, e.g. $\mathbf{x}_{*:t-1}$. 
% This serves to indicate that autoregressive dependencies can have different span including at the extremes the full sequence $\mathbf{x}_{1:t-1}$ and the last value $\mathbf{x}_{t-1}$.
% We have not explicitly listed any cases with multiple latent variables as these generally include a multitude of potential dependency structures.
% We outline notable examples of models with hierarchies of latent variables in section \ref{sec:plvms}.
}
\label{tab:lvm-model-primitives}
\begin{center}
\renewcommand{\arraystretch}{1.1}
\resizebox{0.8\columnwidth}{!}{%
\begin{tabular}{ l l l } 
\toprule
\multicolumn{2}{l}{\textbf{\textsc{Type}}} & \textbf{\textsc{Form}} \\
\midrule
\multicolumn{3}{c}{\textsc{Observation model}} \\
\midrule
\textbf{\textsc{arx}} & Autoregressive on $\mathbf{x}_t$      & $p(\mathbf{x}_t|\mathbf{x}_{1:t-1})$ \\
\textbf{\textsc{loc}} & Local latent variable                 & $p(\mathbf{x}_{t}|\mathbf{z}_{1:t})$ \\
\textbf{\textsc{glb}} & Global latent variable                & $p(\mathbf{x}_{t}|\mathbf{z})$ \\
\midrule
\multicolumn{3}{c}{\textsc{Prior}} \\
\midrule
% \textbf{\textsc{aut}} & Autoregressive on $\mathbf{z}_t$      & $p(\mathbf{z}_t|\mathbf{z}_{1:t-1})$ \\
% \textbf{\textsc{obs}} & Observation dependent                                & $p(\mathbf{z}_t|\mathbf{x}_{1:t-1})$ \\
\textbf{\textsc{arx}} & Autoregressive on $\mathbf{x}_t$      & $p(\mathbf{z}_t|\mathbf{x}_{1:t-1})$ \\
\textbf{\textsc{arz}} & Autoregressive on $\mathbf{z}_t$      & $p(\mathbf{z}_t|\mathbf{z}_{1:t-1})$ \\
\textbf{\textsc{ind}} & Locally independent $\mathbf{z}_t$                 & $p(\mathbf{z}_t)$ \\
\textbf{\textsc{glb}} & Global latent variable                & $p(\mathbf{z})$ \\
\midrule
\multicolumn{3}{c}{\textsc{Inference model}} \\
\midrule
\textbf{\textsc{arz}} & Autoregressive on $\mathbf{z}_t$      & $q(\mathbf{z}_t|\mathbf{z}_{1:t-1})$ \\
\textbf{\textsc{flt}} & Filtering                             & $q(\mathbf{z}_t|\mathbf{x}_{1:t})$ \\
\textbf{\textsc{lsm}} & Local smoothing                       & $q(\mathbf{z}_t|\mathbf{x}_{t-r:t+r})$ \\
\textbf{\textsc{gsm}} & Global smoothing                      & $q(\mathbf{z}_t|\mathbf{x}_{1:T})$ \\
\textbf{\textsc{glb}} & Global latent variable                & $q(\mathbf{z}|\mathbf{x}_{1:T})$ \\
\bottomrule
\end{tabular}
}
\end{center}
\end{table}

\begin{table*}[t]
\caption{
Selected latent variable models classified according the attributes defined throughout section \ref{sec:plvms}.
%The models are sorted according to the publication date of the first version on arXiv which might differ from the year indicated by the citation. 
%This table supplements the classification provided in table \ref{tab:model-taxonomy} which uses binary attributes largely defined for self-supervised methods. 
See table~\ref{tab:lvm-model-primitives} for the probability distributions that correspond to each of the attribute short-hands. \textbf{\textsc{hie}} indicates a hierarchical representation.
%The mathematical expressions of the observation, prior and inference models for the models considered here can be found in table~\ref{tab:lvm-mathematical-expressions}. 
%Figure~\ref{fig:lvm-graphical-models} depicts corresponding graphical model for selected LVMs of this table.
}
\label{tab:lvm-taxonomy}
\begin{center}
\renewcommand{\arraystretch}{1.1}
\resizebox{0.99\textwidth}{!}{% <------ Don't forget this %
\begin{tabular}{ l l | 
c c c | c c c c | c c c c c | c } 
\toprule
\multicolumn{2}{c}{} & 
\multicolumn{3}{c}{\footnotesize\textsc{Observation}} & 
\multicolumn{4}{c}{\footnotesize\textsc{Prior}} & 
\multicolumn{5}{c}{\footnotesize\textsc{Inference}} \\
\textbf{\textsc{model}} & 
\textbf{\textsc{pub. date}} &
\textbf{\textsc{arx}} & 
\textbf{\textsc{loc}} & 
\textbf{\textsc{glb}} &  
\textbf{\textsc{arx}} & 
\textbf{\textsc{arz}} & 
\textbf{\textsc{ind}} & 
\textbf{\textsc{glb}} &  
\textbf{\textsc{arz}} & 
\textbf{\textsc{flt}} & 
\textbf{\textsc{lsm}} & 
\textbf{\textsc{gsm}} & 
\textbf{\textsc{glb}} &
\textbf{\textsc{hie}} \\
\midrule
%                                                              OBSERVATION       |              PRIOR                |                  INFERENCE       
%                                                         ARX      LOC      GLB      ARX      ARZ     IND      GLB      ARZ      FLT      LSM      GSM      GLB

% Conv. DBN \cite{lee_unsupervised_2009}    & 2009 Jan. & \xmark & \cmark & \xmark & \xmark & \xmark & \cmark & \xmark & \xmark & \xmark & \xmark & \xmark & \xmark & \xmark \\
\textbf{VRNN} \footnotesize\cite{chung_recurrent_2015}          & 2015 Jun. & \cmark & \cmark & \xmark & \cmark & \cmark & \xmark & \xmark & \cmark & \cmark & \xmark & \xmark & \xmark & \xmark \\
\textbf{SRNN} \footnotesize\cite{fraccaro_sequential_2016}      & 2016 May  & \cmark & \cmark & \xmark & \cmark & \cmark & \xmark & \xmark & \cmark & \xmark & \xmark & \cmark & \xmark & \xmark \\
\textbf{HMM-VAE} \footnotesize\cite{ebbers_hidden_2017}         & 2017 Mar. & \xmark & \cmark & \xmark & \xmark & \cmark & \xmark & \xmark & \cmark & \cmark & \xmark & \xmark & \xmark & \cmark \\
\textbf{ConvVAE} \footnotesize\cite{hsu_learning_2017}          & 2017 Apr. & \xmark & \xmark & \cmark & \xmark & \xmark & \xmark & \cmark & \xmark & \xmark & \xmark & \cmark & \cmark & \xmark \\
\textbf{FHVAE} \footnotesize\cite{hsu_unsupervised_2017}        & 2017 Sep. & \xmark & \cmark & \cmark & \xmark & \xmark & \cmark & \cmark & \xmark & \xmark & \xmark & \cmark & \cmark & \cmark \\
\textbf{VQ-VAE} (van den Oord et al. 2018)            & 2017 Nov. & \cmark & \cmark & \xmark & \xmark & \xmark & \cmark & \xmark & \xmark & \xmark & \cmark & \xmark & \xmark & \xmark \\
\textbf{BHMM-VAE} \footnotesize\cite{glarner_full_2018}         & 2018 Mar. & \xmark & \cmark & \xmark & \xmark & \cmark & \xmark & \xmark & \cmark & \cmark & \xmark & \xmark & \xmark & \xmark \\
\textbf{STCN} \footnotesize\cite{aksan_stcn_2019}               & 2019 Feb. & \xmark & \cmark & \xmark & \cmark & \xmark & \xmark & \xmark & \xmark & \cmark & \xmark & \xmark & \xmark & \cmark \\
\textbf{FDMM} \footnotesize\cite{khurana_factorial_2019}        & 2019 Oct. & \xmark & \cmark & \cmark & \xmark & \cmark & \xmark & \cmark & \cmark & \cmark & \xmark & \xmark & \cmark & \cmark \\
\textbf{ConvDMM} \footnotesize\cite{khurana_convolutional_2020} & 2020 Jun. & \xmark & \cmark & \xmark & \xmark & \cmark & \xmark & \xmark & \cmark & \xmark & \cmark & \xmark & \xmark & \xmark \\
% WaveNet AE        &  \\
% HMM-VAE           &  \\
% Convolutional DBN &  \\
\bottomrule
\end{tabular}
}
\end{center}
\end{table*}

We next review VAEs applied to speech. We consider the choices of observation, prior and inference models.
% and introduce attributes throughout the text. An overview of these is found in table~\ref{tab:lvm-model-primitives}.
We provide a model taxonomy for selected LVMs in table~\ref{tab:lvm-taxonomy}.
% Supplementing the binary attributes, table~\ref{tab:lvm-mathematical-expressions} explicitly details the observation, prior and inference models of the same selection of LVMs considered in table~\ref{tab:lvm-taxonomy}. 

\subsubsection{Observation models} 
A common choice for the observation model $p(\mathbf{x}|\mathbf{z})$ is to include an autoregressive dependency on the observed variable (\textbf{\textsc{arx}}, table \ref{tab:lvm-model-primitives}) that is, $p(\mathbf{x}_t|\mathbf{x}_{1:t-1}, \cdot)$ where $\cdot$ represents some dependency on the latent variable \cite{chung_recurrent_2015, fraccaro_sequential_2016, oord_neural_2018, oord2018representation}. This allows the latent representation to focus on correlations that cannot easily be predicted from the observed variable at previous time-steps \cite{oord2018representation}. In practice, the dependency on $\mathbf{x}_{1:t-1}$ is often assumed to be Markovian and hence only on $\mathbf{x}_{t-1}$. Another common choice is to depend on a local window $\mathbf{x}_{t-r:t-1}$ where $r>1$ is an integer denoting some receptive field. We will take a dependency on $\mathbf{x}_{1:t-1}$ to mean any one of these choices unless otherwise specified.

While the autoregressive dependency might be important for learning a powerful generative model, it might not benefit the learned latent representations. Specifically encouraging the latent representation to discard correlations across the temporal dimension might degrade the quality of the latent representation.
Furthermore, since such a decoder can perform quite well by simply solving an autoregressive prediction problem, similar to WaveNet \cite{oord_wavenet:_2016}, it can make the model prone to suffer from posterior collapse. This problem arises when the approximate and true posterior distributions collapse into the prior which renders the representations non-informative \cite{bowman_generating_2016, sonderby_ladder_2016}. Notably, posterior collapse is a local minimum of the ELBO since the KL-divergence becomes zero. 
Some works alleviate this problem with tricks like KL-annealing and free bits \cite{bowman_generating_2016, sonderby_ladder_2016, kingma_improved_2016}. 
%KL-annealing down-weighs the KL-divergence in the ELBO (eq.~\ref{eq: lvm likelihood bound recon/kl form (elbo)}) gradually phasing it in during an initial training phase \cite{bowman_generating_2016, sonderby_ladder_2016}. Free bits simply returns a zero gradient for KL-divergences smaller than some number. \cite{kingma_improved_2016}. 
The VQ-VAE uses a quantized latent space that is not susceptible to posterior collapse \textit{per se} \cite{oord_neural_2018}. How to equip LVMs with powerful decoders while avoiding posterior collapse is an open problem.

Some LVMs do not use autoregressive observation models \cite{ebbers_hidden_2017, glarner_full_2018, hsu_unsupervised_2017, hsu_learning_2017, khurana_factorial_2019, khurana_convolutional_2020}. These more closely follow the assumption of \emph{local independence} which states that observed variables are conditionally independent given the local (\textbf{\textsc{loc}}, table \ref{tab:lvm-model-primitives}) and/or global (\textbf{\textsc{glb}}, table \ref{tab:lvm-model-primitives}) latent variables \cite{bartholomew_latent_2011}. 
However, this forces the latent variable to encode details about the observed variable to achieve a good reconstruction. This is opposite to contrastive self-supervised learning which allows models to discard details in $\mathbf{x}_{1:T}$ that do not inform the training objective \cite{baevski2020wav2vec}.

\subsubsection{Priors}
Priors can be said to belong to one or more of four broad categories. See table~\ref{tab:lvm-model-primitives}. 
% Autoregressive dependency on the observed variable (\textbf{\textsc{arx}}, table \ref{tab:lvm-model-primitives}), autoregressive dependency on the latent variable (\textbf{\textsc{arz}}, table \ref{tab:lvm-model-primitives}), locally independent (\textbf{\textsc{ind}}, table \ref{tab:lvm-model-primitives}) and global (\textbf{\textsc{glb}}, table \ref{tab:lvm-model-primitives}).
Priors that are autoregressive on the observed variable (\textbf{\textsc{arx}}, table \ref{tab:lvm-model-primitives}) take the form $p(\mathbf{z}_t|\mathbf{x}_{1:t-1})$. This generally results in a slow-down of the generative process which may be of concern if the use-case is data generation. %It is also interesting to note that this encourages the latent variables to model less dynamic behaviour in their stochastic transitions and instead rely more on the observed variable. 
Priors that are autoregressive on the latent variable (\textbf{\textsc{arz}}, table \ref{tab:lvm-model-primitives}) take the form $p(\mathbf{z}_t|\mathbf{z}_{1:t-1})$ and enable stochastic temporal transitions similar to hidden Markov models but with potentially nonlinear transition functions \cite{chung_recurrent_2015, fraccaro_sequential_2016, khurana_factorial_2019, khurana_convolutional_2020}.
Locally independent priors (\textbf{\textsc{ind}}, table \ref{tab:lvm-model-primitives}) are rarely applied to sequential latent variables since they make the prior latent dynamics independent of the value of previous latent variables. Models that do impose such priors on sequential latents are quite limited in their generative power, unless they learn the prior dynamics post-hoc as done in the VQ-VAE \cite{oord_neural_2018}. 
Global latent variables (\textbf{\textsc{glb}}, table \ref{tab:lvm-model-primitives}) are fundamentally limited in the amount of information they can encode. Hence, models usually use them in combination with another local latent variable, or to encode fixed length input segments \cite{khurana_factorial_2019, hsu_learning_2017, hsu_unsupervised_2017}.

% Many LVMs have priors that do not have autoregressive dependence on the observed variable (table \ref{tab:lvm-model-primitives}). This allows generating the latent sequence in full before generating the observed sequence. It also permits conditioning past observed variables on future latent variables. Despite this, it is usual to let $\mathbf{x}_t$ depend causally on $\mathbf{z}_{1:t}$ in the generative model. To our knowledge, so far no work has examined full (non-causal) dependence of $\mathbf{x}_{1:T}$ on $\mathbf{z}_{1:T}$ during generation.% where, for example, $\mathbf{z}_T$ can influence $\mathbf{x}_1$. 

\subsubsection{Inference models} 
LVMs based on the VAE perform so-called amortized variational inference. Here, a single inference network is used to infer the latent variables of any $\mathbf{x}$. 
For this reason, all inference models covered here are conditioned on the observed sequence in some way. Generally, the inference model can be seen as solving either a filtering or smoothing problem.
In filtering (\textbf{\textsc{flt}}, table \ref{tab:lvm-model-primitives}), the latent variables are assumed to depend only on past and current values of the observed variable, $q(\mathbf{z}_t|\mathbf{x}_{1:t})$ \cite{chung_recurrent_2015, khurana_convolutional_2020}. In global smoothing (\textbf{\textsc{gsm}}, table \ref{tab:lvm-model-primitives}), this causal dependency is replaced with a dependency on all observed values, $q(\mathbf{z}_t|\mathbf{x}_{1:T})$ \cite{fraccaro_sequential_2016, hsu_learning_2017}. 
Smoothing can also be done locally (\textbf{\textsc{lsm}}, table \ref{tab:lvm-model-primitives}), where the latent variables then depend on $\mathbf{x}_{t-r:t+r}$ for some integer $r>0$ \cite{oord_neural_2018}. Compared to self-supervised models that often use transformer encoders it can be hypothesized that global smoothing offers a stronger case than local smoothing and filtering for representation learning.

The inference model may also be used to infer a global latent variable (\textbf{\textsc{glb}}, table \ref{tab:lvm-model-primitives}) that might encode global information about $\mathbf{x}$. 
While it must be included in the prior model it might not be in the observation model, if the model also has a local latent variable. 
Finally, latent variables are often made to depend autoregressively on past inferred values, e.g. $q(\mathbf{z}_t|\mathbf{z}_{1:t-1},\mathbf{x}_{1:t})$ (\textbf{\textsc{arz}}, table \ref{tab:lvm-model-primitives}) \cite{chung_recurrent_2015, fraccaro_sequential_2016}.
% Finally, latent variables are often made to depend autoregressively on past values, $q(\mathbf{z}_t|\mathbf{z}_{1:t-1},\mathbf{x}_{1:t})$ for filtering and $q(\mathbf{z}_t|\mathbf{z}_{1:t-1},\mathbf{x}_{1:T})$ for smoothing (\textbf{\textsc{arz}}, table \ref{tab:lvm-model-primitives}) \cite{chung_recurrent_2015, fraccaro_sequential_2016}.
%Such autoregressive dependencies help the inference model better match the prior dynamics but notably are not employed in models that learn the prior dynamics post-hoc \cite{oord_neural_2018}.

\subsubsection{Multiscale and hierarchical models} Some work has explored using a hierarchy of latent variables (\textbf{\textsc{hie}}, table \ref{tab:lvm-taxonomy}). This allows encoding the inductive bias that speech contains information at different temporal scales by letting the latent variables operate at different temporal scales \cite{hsu_unsupervised_2017}. \citet{khurana_factorial_2019} propose using a temporal latent variable along with a global latent variable. 
Recent work has focused on learning a deeper latent hierarchy with five latent variables \cite{aksan_stcn_2019}. 

\subsubsection{Other work}
% Most recent works learn latent variable models via amortized variational inference as introduced by the VAE framework \cite{kingma_auto-encoding_2014, rezende_stochastic_2014}. 
Before the introduction of the VAE, models such as deep belief networks (DBN, \citealp{hinton_fast_2006}) built from stacks of restricted Boltzmann machines \cite{smolensky_parallel_1987,fischer:13} were popular. \citet{lee_unsupervised_2009} show the feasibility of using a two-layered DBN for discovering acoustic units of speech, while \citet{deng_binary_2010} show that a DBN can learn a binary coding of spectrograms that has higher signal-to-noise ratio than classical vector-quantization techniques for speech coding.
DBNs are however notoriously tricky to optimize requiring the use of expensive MCMC sampling techniques for inference or resort to biased gradient estimates \cite{hinton_practical_2012,fischer:10c}.
Non-neural LVMs for speech representation learning have also been explored
\cite{lee_nonparametric_2012, ondel_variational_2016, heck_feature_2017, jansen_weak_2013}.

\section{Discussion}
\label{sec:mtax}
%Throughout the text we defined several binary attributes which are used to characterize a selection of models in table \ref{tab:model-taxonomy} and \ref{tab:lvm-taxonomy}. We discuss representation learning models based on this overarching model taxonomy below.

%In table \ref{tab:model-taxonomy}, we use those derived from the self-supervised models to characterize a selection of models within the two surveyed categories. To obtain a more detailed understanding of the LVMs, we also derived a set of variable dependencies  presented  in table \ref{tab:lvm-model-primitives}. In table \ref{tab:lvm-taxonomy} we provide a specification of how they apply to selected LVMs. 

% Throughout the text we defined a set of binary model attributes. In table \ref{tab:model-taxonomy}, we use these to characterize a selection of models within the two surveyed categories. These attributes have been derived with a focus on the self-supervised models. As a result, the characterization of the latent variable models is fairly uniform in comparison. Thus, to obtain a more detailed understanding, we also derived a set of dependencies between variables used to specify the graphical models for the LVMs. These are presented in table \ref{tab:lvm-model-primitives} and a specification of how they apply to selected models is given in table \ref{tab:lvm-taxonomy}. Below, we compare models included in this survey using table \ref{tab:model-taxonomy} and \ref{tab:lvm-taxonomy} as a basis for the discussion.

\subsubsection{From global to local} 

In table \ref{tab:model-taxonomy}, we see that work on global representations within self-supervised learning precedes work on local representations. However, we find that the core ideas underlying the recent successes in learning local representation models have also been used for global representation learning; masking \cite{chung2016audio}, context prediction \cite{chung2018speech2vec}, and contrastive training \cite{milde2018unspeech} have been applied in both settings. Furthermore, where work on global representation learning has taken inspiration from Word2vec \cite{mikolov2013distributed}, the techniques used for learning local representations are inspired by contextualized word embeddings \cite{devlin2019bert}. Thus, the gap between these two model classes is largely a product of the developments in related fields and the general increase in computational resources.

\subsubsection{Representations beyond inference} Predictive tasks are commonly used for self-supervised models, but they are not directly compatible with LVM training. 
However, an LVM prior with an autoregressive parameterization, $p(\mathbf{z}_t|\mathbf{z}_{1:t-1})$ or $p(\mathbf{z}_t|\mathbf{x}_{1:t-1})$, can be seen as predictive in the sense that it tries to match the approximate posterior.
%, which conditions on $\mathbf{x}_t$, using only latent dynamics and past observed variables. 
Hence, the prior might be considered for feature extraction.
%although it remains unclear whether a representation extracted from the prior is better suited for downstream tasks than one extracted from the posterior.
\citet{jones2020discrete} examine the importance of the prior in the VQ-VAE and show that the ability of this model to estimate densities $p(\mathbf{x}_{1:T})$ lies solely in its prior. Other work has also explored representations beyond the latent variable such as hidden units of the observation model \cite{khurana_convolutional_2020, chorowski_unsupervised_2019}.

% Hidden units from the observation model have also been used as downstream task features in previous work \cite{khurana_convolutional_2020}. Similarly, \citet{chorowski_unsupervised_2019} explore representations beyond the latent variable.

\subsubsection{Masking and missing data}
Masking may also improve representations learned with VAEs. 
Masking in VAEs has already been explored in the literature in the context of missing data imputation. Here, $\mathbf{x}$ is only partially observed, and often represented as a segmentation into observed and missing parts and a mask $\mathbf{m}$ indicating where the data is missing. The model is then trained to infer the latent variable from the observed data. Reconstruction also deals only with the observed data. Previous work has largely focused on the ability of these models to yield high-quality imputations within the tabular and image data domains, without probing for the effects on the learned latent representation \cite{mattei_miwae_2019, ipsen_not-miwae_2021}. 
The idea of using VAEs to impute missing data was already examined in the seminal paper by \citet{rezende_stochastic_2014}. Here the model was trained with fully observed data and used to impute data in an iterative sampling approach post hoc leaving the learned representations unchanged.

\subsubsection{Evaluating representations} 
Although this review has a primarily methodological focus, we should briefly touch upon evaluation procedures.
Training metrics for self-supervised tasks and the likelihood of LVMs offer little guidance as to the quality of the learned representations \cite{huszar_is_2017}. Thus, a common approach is to evaluate the representations in terms of their usefulness for downstream tasks. Such tasks may be chosen to target specific attributes of the representation (e.g. semantic or speaker information).

The SUPERB benchmark \cite{yang2021superb} gathers multiple tasks grouped into categories such as \emph{recognition}, \emph{detection}, \emph{semantics}, \emph{speaker}, \emph{paralinguistics} and \emph{generation}. 
The recently proposed SLUE benchmark focuses on spoken language understanding \cite{shon2021slue}.
The long-standing zero resource speech challenge (ZeroSpeech) offers a new set of tasks for each edition \cite{versteegh2015zero, dunbar2017zero, dunbar2019zero, dunbar2020zero, dunbar2021zero} usually featuring a minimal-pair ABX task \cite{schatz2013evaluating, schatz2014evaluating}. 
 
Tasks that evaluate representations in terms of speaker-related information include speaker verification \cite{hsu_unsupervised_2017, khurana_factorial_2019, milde2018unspeech}, speaker identification \cite{oord2018representation, jati2019neural, chung2019unsupervised, liu2020non}, dialect classification \cite{khurana_factorial_2019}, emotion recognition \cite{Pascual2019, yang2021superb} and gender classification \cite{lee_unsupervised_2009}. The semantic content of representations are evaluated using tasks such as intent classification \cite{morais2021end, yang2021superb}, slot filling \cite{lai2021semi, yang2021superb}, sentiment analysis \cite{liu2020mockingjay}, question answering \cite{chung2020splat}, named entity recognition \cite{shon2021slue, borgholt2021we, pasad2021use} and speech translation \cite{bansal2017towards, chung2020generative}. Cardiac arrest detection for emergency calls has also been used to evaluate speech representations \cite{borgholt2021we}.
For local representations, phoneme classification is very common \cite{lee_unsupervised_2009, hsu_learning_2017, chorowski_unsupervised_2019, chung2019unsupervised, liu2021tera}.
However, automatic speech recognition has become the \textit{de facto} standard benchmark task \cite{ling2020decoar, chung2020generative, hsu2021hubert}.

\subsubsection{Moving forward} 

Most of the seminal work has focused on improving speech recognition \cite{schneider2019wav2vec, baevski2020wav2vec}. This focus has gained traction over the last couple of years, as computational resources have become more accessible and end-to-end models \cite{graves2006connectionist, chan2016listen} have been established as the dominant approach to speech recognition \cite{gulati2020conformer}. It is important to stress that self-supervised models, such as wav2vec 2.0 \cite{baevski2020wav2vec}, represent a breakthrough, and recent successful approaches build upon this method. That is, deep self-attention models combined with masking \cite{hsu2021hubert, wang2021unispeech, chen2021wavlm}. This development mirrors years of rapid progress in masked language modeling within natural language processing \cite{devlin2019bert,clark_2020_electra} and we expect this to continue for unsupervised neural speech representation learning.

\section{Conclusion}
\label{sec:conc}

We reviewed unsupervised representation learning for speech, focusing on two primary categories: self-supervised methods and probabilistic latent variable models. Inspired by the development of self-supervised learning and the dependency structures of latent variable models, we derived a comprehensive model taxonomy. Finally, we compare and discuss models from the two categories and their respective evaluation procedures.

\bibliography{aaai22}

\begin{thebibliography}{132}
\providecommand{\natexlab}[1]{#1}

\bibitem[{Aksan and Hilliges(2019)}]{aksan_stcn_2019}
Aksan, E.; and Hilliges, O. 2019.
\newblock {STCN}: {Stochastic} {Temporal} {Convolutional} {Networks}.
\newblock \emph{International Conference on Learning Representations (ICLR)}.

\bibitem[{Badino et~al.(2014)Badino, Canevari, Fadiga, and
  Metta}]{badino2014auto}
Badino, L.; Canevari, C.; Fadiga, L.; and Metta, G. 2014.
\newblock An auto-encoder based approach to unsupervised learning of subword
  units.
\newblock \emph{IEEE International Conference on Acoustics, Speech and Signal
  Processing (ICASSP)}.

\bibitem[{Baevski, Auli, and Mohamed(2019)}]{baevski2019effectiveness}
Baevski, A.; Auli, M.; and Mohamed, A. 2019.
\newblock Effectiveness of self-supervised pre-training for speech recognition.
\newblock \emph{arXiv:1911.03912}.

\bibitem[{Baevski et~al.(2021)Baevski, Hsu, Conneau, and
  Auli}]{baevski2021unsupervised}
Baevski, A.; Hsu, W.-N.; Conneau, A.; and Auli, M. 2021.
\newblock Unsupervised Speech Recognition.
\newblock \emph{Neural Information Processing System (NeurIPS)}.

\bibitem[{Baevski, Schneider, and Auli(2019)}]{baevski2019vq}
Baevski, A.; Schneider, S.; and Auli, M. 2019.
\newblock vq-wav2vec: Self-Supervised Learning of Discrete Speech
  Representations.
\newblock \emph{International Conference on Learning Representations (ICLR)}.

\bibitem[{Baevski et~al.(2020)Baevski, Zhou, Mohamed, and
  Auli}]{baevski2020wav2vec}
Baevski, A.; Zhou, Y.; Mohamed, A.; and Auli, M. 2020.
\newblock wav2vec 2.0: A Framework for Self-Supervised Learning of Speech
  Representations.
\newblock \emph{Neural Information Processing System (NeurIPS)}.

\bibitem[{Bansal et~al.(2017)Bansal, Kamper, Lopez, and
  Goldwater}]{bansal2017towards}
Bansal, S.; Kamper, H.; Lopez, A.; and Goldwater, S. 2017.
\newblock Towards speech-to-text translation without speech recognition.
\newblock \emph{European Chapter of the Association for Computational
  Linguistics (EACL)}.

\bibitem[{Bartholomew, Knott, and Moustaki(2011)}]{bartholomew_latent_2011}
Bartholomew, D.~J.; Knott, M.; and Moustaki, I. 2011.
\newblock \emph{Latent variable models and factor analysis: a unified
  approach}.
\newblock Wiley Series in Probability and Statistics. Wiley, 3rd ed edition.

\bibitem[{Bengio, Courville, and Vincent(2013)}]{bengio_representation_2013}
Bengio, Y.; Courville, A.~C.; and Vincent, P. 2013.
\newblock Representation Learning: {A} Review and New Perspectives.
\newblock \emph{{IEEE} Transactions on Pattern Analysis Machine Intelligence},
  35(8): 1798--1828.

\bibitem[{Bengio, L{\'e}onard, and Courville(2013)}]{bengio2013estimating}
Bengio, Y.; L{\'e}onard, N.; and Courville, A. 2013.
\newblock Estimating or propagating gradients through stochastic neurons for
  conditional computation.
\newblock \emph{arXiv:1308.3432}.

\bibitem[{Bhati et~al.(2021{\natexlab{a}})Bhati, Villalba, {\.Z}elasko,
  Moro-Velazquez, and Dehak}]{bhati2021segmental}
Bhati, S.; Villalba, J.; {\.Z}elasko, P.; Moro-Velazquez, L.; and Dehak, N.
  2021{\natexlab{a}}.
\newblock Segmental Contrastive Predictive Coding for Unsupervised Word
  Segmentation.
\newblock \emph{arXiv:2106.02170}.

\bibitem[{Bhati et~al.(2021{\natexlab{b}})Bhati, Villalba, {\.Z}elasko,
  Moro-Velazquez, and Dehak}]{bhati2021unsupervised}
Bhati, S.; Villalba, J.; {\.Z}elasko, P.; Moro-Velazquez, L.; and Dehak, N.
  2021{\natexlab{b}}.
\newblock Unsupervised Speech Segmentation and Variable Rate Representation
  Learning using Segmental Contrastive Predictive Coding.
\newblock \emph{arXiv:2110.02345}.

\bibitem[{Borgholt et~al.(2021{\natexlab{a}})Borgholt, Havtorn, Abdou, Edin,
  Maal{\o}e, S{\o}gaard, and Igel}]{borgholt2021we}
Borgholt, L.; Havtorn, J.~D.; Abdou, M.; Edin, J.; Maal{\o}e, L.; S{\o}gaard,
  A.; and Igel, C. 2021{\natexlab{a}}.
\newblock Do We Still Need Automatic Speech Recognition for Spoken Language
  Understanding?
\newblock \emph{arXiv:2111.14842}.

\bibitem[{Borgholt et~al.(2021{\natexlab{b}})Borgholt, Tax, Havtorn, Maal{\o}e,
  and Igel}]{borgholt2021scaling}
Borgholt, L.; Tax, T.~M.; Havtorn, J.~D.; Maal{\o}e, L.; and Igel, C.
  2021{\natexlab{b}}.
\newblock On Scaling Contrastive Representations for Low-Resource Speech
  Recognition.
\newblock \emph{IEEE International Conference on Acoustics, Speech and Signal
  Processing (ICASSP)}.

\bibitem[{Bowman et~al.(2016)Bowman, Vilnis, Vinyals, Dai, Józefowicz, and
  Bengio}]{bowman_generating_2016}
Bowman, S.~R.; Vilnis, L.; Vinyals, O.; Dai, A.~M.; Józefowicz, R.; and
  Bengio, S. 2016.
\newblock Generating {Sentences} from a {Continuous} {Space}.
\newblock \emph{{SIGNLL} {Conference} on {Computational} {Natural} {Language}
  {Learning} ({CoNLL})}.

\bibitem[{Burton, Shore, and Buck(1983)}]{burton_generalization_1983}
Burton, D.~K.; Shore, J.~E.; and Buck, J.~T. 1983.
\newblock A generalization of isolated word recognition using vector
  quantization.
\newblock \emph{IEEE International Conference on Acoustics, Speech and Signal
  Processing (ICASSP)}.

\bibitem[{Chan et~al.(2016)Chan, Jaitly, Le, and Vinyals}]{chan2016listen}
Chan, W.; Jaitly, N.; Le, Q.; and Vinyals, O. 2016.
\newblock Listen, attend and spell: A neural network for large vocabulary
  conversational speech recognition.
\newblock \emph{IEEE International Conference on Acoustics, Speech and Signal
  Processing (ICASSP)}.

\bibitem[{Chen et~al.(2021)Chen, Wang, Chen, Wu, Liu, Chen, Li, Kanda,
  Yoshioka, Xiao et~al.}]{chen2021wavlm}
Chen, S.; Wang, C.; Chen, Z.; Wu, Y.; Liu, S.; Chen, Z.; Li, J.; Kanda, N.;
  Yoshioka, T.; Xiao, X.; et~al. 2021.
\newblock WavLM: Large-Scale Self-Supervised Pre-Training for Full Stack Speech
  Processing.
\newblock \emph{arXiv:2110.13900}.

\bibitem[{Chi et~al.(2021)Chi, Chung, Wu, Hsieh, Chen, Li, and
  Lee}]{chi2021audio}
Chi, P.-H.; Chung, P.-H.; Wu, T.-H.; Hsieh, C.-C.; Chen, Y.-H.; Li, S.-W.; and
  Lee, H.-y. 2021.
\newblock Audio albert: A lite bert for self-supervised learning of audio
  representation.
\newblock \emph{IEEE Spoken Language Technology Workshop (SLT)}.

\bibitem[{Chorowski et~al.(2019{\natexlab{a}})Chorowski, Chen, Marxer, Dolfing,
  {\L}a{\'n}cucki, Sanchez, Alum{\"a}e, and
  Laurent}]{chorowski2019unsupervised}
Chorowski, J.; Chen, N.; Marxer, R.; Dolfing, H.; {\L}a{\'n}cucki, A.; Sanchez,
  G.; Alum{\"a}e, T.; and Laurent, A. 2019{\natexlab{a}}.
\newblock Unsupervised neural segmentation and clustering for unit discovery in
  sequential data.
\newblock \emph{NeurIPS workshop: Perception as generative reasoning}.

\bibitem[{Chorowski et~al.(2019{\natexlab{b}})Chorowski, Weiss, Bengio, and
  van~den Oord}]{chorowski_unsupervised_2019}
Chorowski, J.; Weiss, R.~J.; Bengio, S.; and van~den Oord, A.
  2019{\natexlab{b}}.
\newblock Unsupervised speech representation learning using {WaveNet}
  autoencoders.
\newblock \emph{IEEE/ACM Transactions on Audio, Speech, and Language
  Processing}.

\bibitem[{Chung et~al.(2015)Chung, Kastner, Dinh, Goel, Courville, and
  Bengio}]{chung_recurrent_2015}
Chung, J.; Kastner, K.; Dinh, L.; Goel, K.; Courville, A.~C.; and Bengio, Y.
  2015.
\newblock A {Recurrent} {Latent} {Variable} {Model} for {Sequential} {Data}.
\newblock \emph{Neural Information Processing System (NeurIPS)}.

\bibitem[{Chung and Glass(2017)}]{chung2017learning}
Chung, Y.-A.; and Glass, J. 2017.
\newblock Learning word embeddings from speech.
\newblock \emph{NeurIPS ML4Audio Workshop}.

\bibitem[{Chung and Glass(2018)}]{chung2018speech2vec}
Chung, Y.-A.; and Glass, J. 2018.
\newblock Speech2vec: A sequence-to-sequence framework for learning word
  embeddings from speech.
\newblock \emph{INTERSPEECH}.

\bibitem[{Chung and Glass(2020{\natexlab{a}})}]{chung2020generative}
Chung, Y.-A.; and Glass, J. 2020{\natexlab{a}}.
\newblock Generative pre-training for speech with autoregressive predictive
  coding.
\newblock \emph{IEEE International Conference on Acoustics, Speech and Signal
  Processing (ICASSP)}.

\bibitem[{Chung and Glass(2020{\natexlab{b}})}]{chung2020improved}
Chung, Y.-A.; and Glass, J. 2020{\natexlab{b}}.
\newblock Improved speech representations with multi-target autoregressive
  predictive coding.
\newblock \emph{Annual Meeting of the Association for Computational Linguistics
  (ACL)}.

\bibitem[{Chung et~al.(2019)Chung, Hsu, Tang, and
  Glass}]{chung2019unsupervised}
Chung, Y.-A.; Hsu, W.-N.; Tang, H.; and Glass, J.~R. 2019.
\newblock An Unsupervised Autoregressive Model for Speech Representation
  Learning.
\newblock \emph{INTERSPEECH}.

\bibitem[{Chung, Tang, and Glass(2020)}]{chung2020vector}
Chung, Y.-A.; Tang, H.; and Glass, J. 2020.
\newblock Vector-Quantized Autoregressive Predictive Coding.
\newblock \emph{INTERSPEECH}.

\bibitem[{Chung et~al.(2016)Chung, Wu, Shen, Lee, and Lee}]{chung2016audio}
Chung, Y.-A.; Wu, C.-C.; Shen, C.-H.; Lee, H.-Y.; and Lee, L.-S. 2016.
\newblock Audio word2vec: Unsupervised learning of audio segment
  representations using sequence-to-sequence autoencoder.
\newblock \emph{INTERSPEECH}.

\bibitem[{Chung, Zhu, and Zeng(2020)}]{chung2020splat}
Chung, Y.-A.; Zhu, C.; and Zeng, M. 2020.
\newblock SPLAT: Speech-Language Joint Pre-Training for Spoken Language
  Understanding.
\newblock \emph{arXiv:2010.02295}.

\bibitem[{Clark et~al.(2020)Clark, Luong, Le, and Manning}]{clark_2020_electra}
Clark, K.; Luong, M.; Le, Q.~V.; and Manning, C.~D. 2020.
\newblock {ELECTRA:} Pre-training Text Encoders as Discriminators Rather Than
  Generators.
\newblock \emph{International Conference on Learning Representations (ICLR)}.

\bibitem[{Conneau et~al.(2020)Conneau, Baevski, Collobert, Mohamed, and
  Auli}]{conneau2020unsupervised}
Conneau, A.; Baevski, A.; Collobert, R.; Mohamed, A.; and Auli, M. 2020.
\newblock Unsupervised cross-lingual representation learning for speech
  recognition.
\newblock \emph{INTERSPEECH}.

\bibitem[{Cuervo et~al.(2021)Cuervo, Grabias, Chorowski, Ciesielski,
  {\L}a{\'n}cucki, Rychlikowski, and Marxer}]{cuervo2021contrastive}
Cuervo, S.; Grabias, M.; Chorowski, J.; Ciesielski, G.; {\L}a{\'n}cucki, A.;
  Rychlikowski, P.; and Marxer, R. 2021.
\newblock Contrastive prediction strategies for unsupervised segmentation and
  categorization of phonemes and words.
\newblock \emph{arXiv:2110.15909}.

\bibitem[{Deng et~al.(2009)Deng, Dong, Socher, Li, Li, and
  Fei-Fei}]{deng2009imagenet}
Deng, J.; Dong, W.; Socher, R.; Li, L.-J.; Li, K.; and Fei-Fei, L. 2009.
\newblock Imagenet: A large-scale hierarchical image database.
\newblock \emph{IEEE Conference on Computer Vision and Pattern Recognition
  (CVPR)}.

\bibitem[{Deng et~al.(2010)Deng, Seltzer, Yu, Acero, Mohamed, and
  Hinton}]{deng_binary_2010}
Deng, L.; Seltzer, M.~L.; Yu, D.; Acero, A.; Mohamed, A.; and Hinton, G.~E.
  2010.
\newblock Binary coding of speech spectrograms using a deep auto-encoder.
\newblock \emph{INTERSPEECH}.

\bibitem[{Devlin et~al.(2019)Devlin, Chang, Lee, and
  Toutanova}]{devlin2019bert}
Devlin, J.; Chang, M.-W.; Lee, K.; and Toutanova, K. 2019.
\newblock {BERT}: Pre-training of Deep Bidirectional Transformers for Language
  Understanding.
\newblock \emph{North American Chapter of the Association for Computational
  Linguistics (NAACL)}.

\bibitem[{Dieleman et~al.(2021)Dieleman, Nash, Engel, and
  Simonyan}]{dieleman_variable-rate_2021}
Dieleman, S.; Nash, C.; Engel, J.; and Simonyan, K. 2021.
\newblock Variable-rate discrete representation learning.
\newblock \emph{arXiv:2103.06089}.

\bibitem[{Doersch, Gupta, and Efros(2015)}]{doersch2015unsupervised}
Doersch, C.; Gupta, A.; and Efros, A.~A. 2015.
\newblock Unsupervised visual representation learning by context prediction.
\newblock \emph{IEEE Conference on Computer Vision and Pattern Recognition
  (CVPR)}.

\bibitem[{Dunbar et~al.(2019)Dunbar, Algayres, Karadayi, Bernard, Benjumea,
  Cao, Miskic, Dugrain, Ondel, Black et~al.}]{dunbar2019zero}
Dunbar, E.; Algayres, R.; Karadayi, J.; Bernard, M.; Benjumea, J.; Cao, X.-N.;
  Miskic, L.; Dugrain, C.; Ondel, L.; Black, A.; et~al. 2019.
\newblock The {Zero Resource Speech Challenge} 2019: {TTS} without {T}.
\newblock \emph{INTERSPEECH}.

\bibitem[{Dunbar et~al.(2021)Dunbar, Bernard, Hamilakis, Nguyen, de~Seyssel,
  Roz{\'e}, Rivi{\`e}re, Kharitonov, and Dupoux}]{dunbar2021zero}
Dunbar, E.; Bernard, M.; Hamilakis, N.; Nguyen, T.; de~Seyssel, M.; Roz{\'e},
  P.; Rivi{\`e}re, M.; Kharitonov, E.; and Dupoux, E. 2021.
\newblock The {Zero Resource Speech Challenge} 2021: Spoken language modelling.
\newblock \emph{IEEE Transactions on Pattern Analysis and Machine
  Intelligence}.

\bibitem[{Dunbar et~al.(2017)Dunbar, Cao, Benjumea, Karadayi, Bernard,
  Besacier, Anguera, and Dupoux}]{dunbar2017zero}
Dunbar, E.; Cao, X.~N.; Benjumea, J.; Karadayi, J.; Bernard, M.; Besacier, L.;
  Anguera, X.; and Dupoux, E. 2017.
\newblock The {Zero Resource Speech Challenge} 2017.
\newblock \emph{IEEE Workshop on Automatic Speech Recognition and Understanding
  (ASRU)}.

\bibitem[{Dunbar et~al.(2020)Dunbar, Karadayi, Bernard, Cao, Algayres, Ondel,
  Besacier, Sakti, and Dupoux}]{dunbar2020zero}
Dunbar, E.; Karadayi, J.; Bernard, M.; Cao, X.-N.; Algayres, R.; Ondel, L.;
  Besacier, L.; Sakti, S.; and Dupoux, E. 2020.
\newblock The {Zero Resource Speech Challenge} 2020: Discovering discrete
  subword and word units.
\newblock \emph{INTERSPEECH}.

\bibitem[{Ebbers et~al.(2017)Ebbers, Heymann, Drude, Glarner, Haeb-Umbach, and
  Raj}]{ebbers_hidden_2017}
Ebbers, J.; Heymann, J.; Drude, L.; Glarner, T.; Haeb-Umbach, R.; and Raj, B.
  2017.
\newblock Hidden {Markov} {Model} {Variational} {Autoencoder} for {Acoustic}
  {Unit} {Discovery}.
\newblock \emph{INTERSPEECH}.

\bibitem[{Fischer and Igel(2011)}]{fischer:10c}
Fischer, A.; and Igel, C. 2011.
\newblock Bounding the Bias of Contrastive Divergence Learning.
\newblock \emph{Neural Computation}, 23: 664--673.

\bibitem[{Fischer and Igel(2014)}]{fischer:13}
Fischer, A.; and Igel, C. 2014.
\newblock Training Restricted {B}oltzmann Machines: {An} Introduction.
\newblock \emph{Pattern Recognition}, 47: 25--39.

\bibitem[{Fraccaro et~al.(2016)Fraccaro, Sønderby, Paquet, and
  Winther}]{fraccaro_sequential_2016}
Fraccaro, M.; Sønderby, S.~K.; Paquet, U.; and Winther, O. 2016.
\newblock Sequential {Neural} {Models} with {Stochastic} {Layers}.
\newblock \emph{Neural Information Processing System (NeurIPS)}.

\bibitem[{Garofolo(1993)}]{garofolo_timit_1993}
Garofolo, J.~S. 1993.
\newblock {TIMIT} {Acoustic}-{Phonetic} {Continuous} {Speech} {Corpus}
  {LDC93S1}.
\newblock Web Download.

\bibitem[{Glarner et~al.(2018)Glarner, Hanebrink, Ebbers, and
  Haeb-Umbach}]{glarner_full_2018}
Glarner, T.; Hanebrink, P.; Ebbers, J.; and Haeb-Umbach, R. 2018.
\newblock Full {Bayesian} Hidden {Markov} Model Variational Autoencoder for
  Acoustic Unit Discovery.
\newblock \emph{INTERSPEECH}.

\bibitem[{Graves et~al.(2006)Graves, Fern{\'a}ndez, Gomez, and
  Schmidhuber}]{graves2006connectionist}
Graves, A.; Fern{\'a}ndez, S.; Gomez, F.; and Schmidhuber, J. 2006.
\newblock Connectionist temporal classification: labelling unsegmented sequence
  data with recurrent neural networks.
\newblock \emph{International Conference on Machine Learning (ICML)}.

\bibitem[{Gulati et~al.(2020)Gulati, Qin, Chiu, Parmar, Zhang, Yu, Han, Wang,
  Zhang, Wu et~al.}]{gulati2020conformer}
Gulati, A.; Qin, J.; Chiu, C.-C.; Parmar, N.; Zhang, Y.; Yu, J.; Han, W.; Wang,
  S.; Zhang, Z.; Wu, Y.; et~al. 2020.
\newblock Conformer: Convolution-augmented Transformer for Speech Recognition.
\newblock \emph{INTERSPEECH}.

\bibitem[{Gutmann and Hyv{\"a}rinen(2010)}]{gutmann2010noise}
Gutmann, M.; and Hyv{\"a}rinen, A. 2010.
\newblock Noise-contrastive estimation: A new estimation principle for
  unnormalized statistical models.
\newblock \emph{International Conference on Artificial Intelligence and
  Statistics (AISTATS)}.

\bibitem[{He et~al.(2016)He, Zhang, Ren, and Sun}]{he2016deep}
He, K.; Zhang, X.; Ren, S.; and Sun, J. 2016.
\newblock Deep residual learning for image recognition.
\newblock \emph{IEEE Conference on Computer Vision and Pattern Recognition
  (CVPR)}.

\bibitem[{Heck, Sakti, and Nakamura(2017)}]{heck_feature_2017}
Heck, M.; Sakti, S.; and Nakamura, S. 2017.
\newblock Feature optimized {DPGMM} clustering for unsupervised subword
  modeling: {A} contribution to {ZeroSpeech} 2017.
\newblock \emph{2017 {IEEE} Automatic Speech Recognition and Understanding
  Workshop, {ASRU} 2017}.

\bibitem[{Hinton(2012)}]{hinton_practical_2012}
Hinton, G.~E. 2012.
\newblock A Practical Guide to Training Restricted Boltzmann Machines.
\newblock In Montavon, G.; Orr, G.~B.; and M{\"u}ller, K.-R., eds.,
  \emph{Neural Networks: Tricks of the Trade: Second Edition}, 599--619.
  Springer.

\bibitem[{Hinton, Osindero, and Teh(2006)}]{hinton_fast_2006}
Hinton, G.~E.; Osindero, S.; and Teh, Y.~W. 2006.
\newblock A Fast Learning Algorithm for Deep Belief Nets.
\newblock \emph{Neural Computation}, 18(7): 1527--1554.

\bibitem[{Holzenberger et~al.(2018)Holzenberger, Du, Karadayi, Riad, and
  Dupoux}]{holzenberger2018learning}
Holzenberger, N.; Du, M.; Karadayi, J.; Riad, R.; and Dupoux, E. 2018.
\newblock Learning word embeddings: Unsupervised methods for fixed-size
  representations of variable-length speech segments.
\newblock \emph{INTERSPEECH}.

\bibitem[{Hsu et~al.(2021)Hsu, Bolte, Tsai, Lakhotia, Salakhutdinov, and
  Mohamed}]{hsu2021hubert}
Hsu, W.-N.; Bolte, B.; Tsai, Y.-H.~H.; Lakhotia, K.; Salakhutdinov, R.; and
  Mohamed, A. 2021.
\newblock HuBERT: Self-Supervised Speech Representation Learning by Masked
  Prediction of Hidden Units.
\newblock \emph{arXiv:2106.07447}.

\bibitem[{Hsu, Zhang, and Glass(2017{\natexlab{a}})}]{hsu_learning_2017}
Hsu, W.-N.; Zhang, Y.; and Glass, J. 2017{\natexlab{a}}.
\newblock Learning {Latent} {Representations} for {Speech} {Generation} and
  {Transformation}.
\newblock \emph{INTERSPEECH}.

\bibitem[{Hsu, Zhang, and Glass(2017{\natexlab{b}})}]{hsu_unsupervised_2017}
Hsu, W.-N.; Zhang, Y.; and Glass, J. 2017{\natexlab{b}}.
\newblock Unsupervised {Learning} of {Disentangled} and {Interpretable}
  {Representations} from {Sequential} {Data}.
\newblock \emph{Neural Information Processing System (NeurIPS)}.

\bibitem[{Huszár(2017)}]{huszar_is_2017}
Huszár, F. 2017.
\newblock Is {Maximum} {Likelihood} {Useful} for {Representation} {Learning}?
\newblock \emph{inFERENCe}.

\bibitem[{Ipsen, Mattei, and Frellsen(2021)}]{ipsen_not-miwae_2021}
Ipsen, N.~B.; Mattei, P.-A.; and Frellsen, J. 2021.
\newblock not-{MIWAE}: {Deep} generative modelling with missing not at random
  data.
\newblock \emph{International Conference on Learning Representations (ICLR)}.

\bibitem[{Jacobs, Matusevych, and Kamper(2021)}]{jacobs2021acoustic}
Jacobs, C.; Matusevych, Y.; and Kamper, H. 2021.
\newblock Acoustic word embeddings for zero-resource languages using
  self-supervised contrastive learning and multilingual adaptation.
\newblock \emph{IEEE Spoken Language Technology Workshop (SLT)}.

\bibitem[{Jang, Gu, and Poole(2017)}]{jang2016categorical}
Jang, E.; Gu, S.; and Poole, B. 2017.
\newblock Categorical Reparameterization with Gumbel-Softmax.
\newblock \emph{International Conference on Learning Representations (ICLR)}.

\bibitem[{Jansen et~al.(2018)Jansen, Plakal, Pandya, Ellis, Hershey, Liu,
  Moore, and Saurous}]{jansen2018unsupervised}
Jansen, A.; Plakal, M.; Pandya, R.; Ellis, D.~P.; Hershey, S.; Liu, J.; Moore,
  R.~C.; and Saurous, R.~A. 2018.
\newblock Unsupervised learning of semantic audio representations.
\newblock \emph{IEEE International Conference on Acoustics, Speech and Signal
  Processing (ICASSP)}.

\bibitem[{Jansen, Thomas, and Hermansky(2013)}]{jansen_weak_2013}
Jansen, A.; Thomas, S.; and Hermansky, H. 2013.
\newblock Weak top-down constraints for unsupervised acoustic model training.
\newblock \emph{IEEE International Conference on Acoustics, Speech and Signal
  Processing (ICASSP)}.

\bibitem[{Jati and Georgiou(2019)}]{jati2019neural}
Jati, A.; and Georgiou, P. 2019.
\newblock Neural predictive coding using convolutional neural networks toward
  unsupervised learning of speaker characteristics.
\newblock \emph{IEEE/ACM Transactions on Audio, Speech, and Language
  Processing}, 27.

\bibitem[{Jati and Georgiou(2017)}]{jati2017speaker2vec}
Jati, A.; and Georgiou, P.~G. 2017.
\newblock {Speaker2Vec}: Unsupervised Learning and Adaptation of a Speaker
  Manifold Using Deep Neural Networks with an Evaluation on Speaker
  Segmentation.
\newblock \emph{INTERSPEECH}.

\bibitem[{Jiang et~al.(2019)Jiang, Lei, Li, Luo, Hu, Zou, and
  Li}]{jiang2019improving}
Jiang, D.; Lei, X.; Li, W.; Luo, N.; Hu, Y.; Zou, W.; and Li, X. 2019.
\newblock Improving transformer-based speech recognition using unsupervised
  pre-training.
\newblock \emph{arXiv:1910.09932}.

\bibitem[{Jiang et~al.(2021)Jiang, Li, Zhang, Cao, Luo, Han, Zou, Han, and
  Li}]{jiang2021further}
Jiang, D.; Li, W.; Zhang, R.; Cao, M.; Luo, N.; Han, Y.; Zou, W.; Han, K.; and
  Li, X. 2021.
\newblock A Further Study of Unsupervised Pretraining for Transformer Based
  Speech Recognition.
\newblock \emph{IEEE International Conference on Acoustics, Speech and Signal
  Processing (ICASSP)}.

\bibitem[{Jones and Moore(2020)}]{jones2020discrete}
Jones, H.~T.; and Moore, J. 2020.
\newblock Is the Discrete {VAE}’s Power Stuck in its Prior?
\newblock \emph{''I Can't Believe It's Not Better!'' NeurIPS 2020 workshop}.

\bibitem[{Jordan et~al.(1999)Jordan, Ghahramani, Jaakkola, and
  Saul}]{jordan_introduction_1999}
Jordan, M.~I.; Ghahramani, Z.; Jaakkola, T.~S.; and Saul, L.~K. 1999.
\newblock An Introduction to Variational Methods for Graphical Models.
\newblock \emph{Machine Learning}, 37(2): 183--233.

\bibitem[{Kamper(2019)}]{kamper2019truly}
Kamper, H. 2019.
\newblock Truly unsupervised acoustic word embeddings using weak top-down
  constraints in encoder-decoder models.
\newblock \emph{IEEE International Conference on Acoustics, Speech and Signal
  Processing (ICASSP)}.

\bibitem[{Kamper et~al.(2015)Kamper, Elsner, Jansen, and
  Goldwater}]{kamper2015unsupervised}
Kamper, H.; Elsner, M.; Jansen, A.; and Goldwater, S. 2015.
\newblock Unsupervised neural network based feature extraction using weak
  top-down constraints.
\newblock \emph{IEEE International Conference on Acoustics, Speech and Signal
  Processing (ICASSP)}.

\bibitem[{Kamper and van Niekerk(2021)}]{kamper2020towards}
Kamper, H.; and van Niekerk, B. 2021.
\newblock Towards unsupervised phone and word segmentation using
  self-supervised vector-quantized neural networks.
\newblock \emph{INTERSPEECH}.

\bibitem[{Kawakami et~al.(2020)Kawakami, Wang, Dyer, Blunsom, and van~den
  Oord}]{kawakami2020learning}
Kawakami, K.; Wang, L.; Dyer, C.; Blunsom, P.; and van~den Oord, A. 2020.
\newblock Learning Robust and Multilingual Speech Representations.
\newblock \emph{Conference on Empirical Methods in Natural Language Processing
  (EMNLP)}.

\bibitem[{Khurana et~al.(2019)Khurana, Joty, Ali, and
  Glass}]{khurana_factorial_2019}
Khurana, S.; Joty, S.~R.; Ali, A.; and Glass, J. 2019.
\newblock A {Factorial} {Deep} {Markov} {Model} for {Unsupervised}
  {Disentangled} {Representation} {Learning} from {Speech}.
\newblock \emph{IEEE International Conference on Acoustics, Speech and Signal
  Processing (ICASSP)}.

\bibitem[{Khurana, Laurent, and Glass(2021)}]{khurana2021magic}
Khurana, S.; Laurent, A.; and Glass, J. 2021.
\newblock Magic dust for cross-lingual adaptation of monolingual wav2vec-2.0.
\newblock \emph{arXiv:2110.03560}.

\bibitem[{Khurana et~al.(2020)Khurana, Laurent, Hsu, Chorowski, Lancucki,
  Marxer, and Glass}]{khurana_convolutional_2020}
Khurana, S.; Laurent, A.; Hsu, W.-N.; Chorowski, J.; Lancucki, A.; Marxer, R.;
  and Glass, J. 2020.
\newblock A {Convolutional} {Deep} {Markov} {Model} for {Unsupervised} {Speech}
  {Representation} {Learning}.
\newblock \emph{INTERSPEECH}.

\bibitem[{Kingma et~al.(2016)Kingma, Salimans, Jozefowicz, Chen, Sutskever, and
  Welling}]{kingma_improved_2016}
Kingma, D.~P.; Salimans, T.; Jozefowicz, R.; Chen, X.; Sutskever, I.; and
  Welling, M. 2016.
\newblock Improved Variational Inference with Inverse Autoregressive Flow.
\newblock \emph{Neural Information Processing System (NeurIPS)}.

\bibitem[{Kingma and Welling(2014)}]{kingma_auto-encoding_2014}
Kingma, D.~P.; and Welling, M. 2014.
\newblock Auto-{Encoding} {Variational} {Bayes}.
\newblock \emph{International Conference on Learning Representations (ICLR)}.

\bibitem[{Kramer(1991)}]{kramer1991nonlinear}
Kramer, M.~A. 1991.
\newblock Nonlinear principal component analysis using autoassociative neural
  networks.
\newblock \emph{AIChE Journal}, 37(2): 233--243.

\bibitem[{Kreuk, Keshet, and Adi(2020)}]{kreuk2020self}
Kreuk, F.; Keshet, J.; and Adi, Y. 2020.
\newblock Self-Supervised Contrastive Learning for Unsupervised Phoneme
  Segmentation.
\newblock \emph{INTERSPEECH}.

\bibitem[{Lai et~al.(2021)Lai, Chuang, Lee, Li, and Glass}]{lai2021semi}
Lai, C.-I.; Chuang, Y.-S.; Lee, H.-Y.; Li, S.-W.; and Glass, J. 2021.
\newblock Semi-supervised spoken language understanding via self-supervised
  speech and language model pretraining.
\newblock \emph{IEEE International Conference on Acoustics, Speech and Signal
  Processing (ICASSP)}.

\bibitem[{Lee and Glass(2012)}]{lee_nonparametric_2012}
Lee, C.-y.; and Glass, J. 2012.
\newblock A {Nonparametric} {Bayesian} {Approach} to {Acoustic} {Model}
  {Discovery}.
\newblock \emph{Annual Meeting of the Association for Computational Linguistics
  (ACL)}.

\bibitem[{Lee et~al.(2009)Lee, Largman, Pham, and Ng}]{lee_unsupervised_2009}
Lee, H.; Largman, Y.; Pham, P.; and Ng, A.~Y. 2009.
\newblock Unsupervised feature learning for audio classification using
  convolutional deep belief networks.
\newblock \emph{Neural Information Processing System (NeurIPS)}.

\bibitem[{Levin et~al.(2013)Levin, Henry, Jansen, and Livescu}]{levin2013fixed}
Levin, K.; Henry, K.; Jansen, A.; and Livescu, K. 2013.
\newblock Fixed-dimensional acoustic embeddings of variable-length segments in
  low-resource settings.
\newblock \emph{IEEE Workshop on Automatic Speech Recognition and Understanding
  (ASRU)}.

\bibitem[{Ling and Liu(2020)}]{ling2020decoar}
Ling, S.; and Liu, Y. 2020.
\newblock Decoar 2.0: Deep contextualized acoustic representations with vector
  quantization.
\newblock \emph{arXiv:2012.06659}.

\bibitem[{Ling et~al.(2020)Ling, Liu, Salazar, and Kirchhoff}]{ling2020deep}
Ling, S.; Liu, Y.; Salazar, J.; and Kirchhoff, K. 2020.
\newblock Deep contextualized acoustic representations for semi-supervised
  speech recognition.
\newblock \emph{IEEE International Conference on Acoustics, Speech and Signal
  Processing (ICASSP)}.

\bibitem[{Liu, Chung, and Glass(2021)}]{liu2020non}
Liu, A.~H.; Chung, Y.-A.; and Glass, J. 2021.
\newblock Non-autoregressive predictive coding for learning speech
  representations from local dependencies.
\newblock \emph{INTERSPEECH}.

\bibitem[{Liu, Li, and Lee(2021)}]{liu2021tera}
Liu, A.~T.; Li, S.-W.; and Lee, H.-y. 2021.
\newblock Tera: Self-supervised learning of transformer encoder representation
  for speech.
\newblock \emph{IEEE/ACM Transactions on Audio, Speech, and Language
  Processing}, 29: 2351--2366.

\bibitem[{Liu et~al.(2020)Liu, Yang, Chi, Hsu, and Lee}]{liu2020mockingjay}
Liu, A.~T.; Yang, S.-w.; Chi, P.-H.; Hsu, P.-c.; and Lee, H.-y. 2020.
\newblock Mockingjay: Unsupervised speech representation learning with deep
  bidirectional transformer encoders.
\newblock \emph{IEEE International Conference on Acoustics, Speech and Signal
  Processing (ICASSP)}.

\bibitem[{Maddison, Mnih, and Teh(2017)}]{maddison2016concrete}
Maddison, C.~J.; Mnih, A.; and Teh, Y.~W. 2017.
\newblock The concrete distribution: A continuous relaxation of discrete random
  variables.
\newblock \emph{International Conference on Learning Representations (ICLR)}.

\bibitem[{Mattei and Frellsen(2019)}]{mattei_miwae_2019}
Mattei, P.-A.; and Frellsen, J. 2019.
\newblock {MIWAE}: {Deep} generative modelling and imputation of incomplete
  data sets.
\newblock \emph{International Conference on Machine Learning (ICML)}.

\bibitem[{Michel et~al.(2017)Michel, Rasanen, Thiolli{\`e}re, and
  Dupoux}]{michel2017blind}
Michel, P.; Rasanen, O.; Thiolli{\`e}re, R.; and Dupoux, E. 2017.
\newblock Blind Phoneme Segmentation With Temporal Prediction Errors.
\newblock \emph{ACL Student Research Workshop}.

\bibitem[{Mikolov et~al.(2010)Mikolov, Karafi{\'a}t, Burget, Cernock{\`y}, and
  Khudanpur}]{mikolov2010recurrent}
Mikolov, T.; Karafi{\'a}t, M.; Burget, L.; Cernock{\`y}, J.; and Khudanpur, S.
  2010.
\newblock Recurrent neural network based language model.
\newblock \emph{INTERSPEECH}.

\bibitem[{Mikolov et~al.(2013)Mikolov, Sutskever, Chen, Corrado, and
  Dean}]{mikolov2013distributed}
Mikolov, T.; Sutskever, I.; Chen, K.; Corrado, G.~S.; and Dean, J. 2013.
\newblock Distributed representations of words and phrases and their
  compositionality.
\newblock \emph{Neural Information Processing System (NeurIPS)}.

\bibitem[{Milde and Biemann(2018)}]{milde2018unspeech}
Milde, B.; and Biemann, C. 2018.
\newblock Unspeech: Unsupervised speech context embeddings.
\newblock \emph{INTERSPEECH}.

\bibitem[{Mohamed et~al.(2020)Mohamed, Rosca, Figurnov, and
  Mnih}]{mohamed_monte_2019}
Mohamed, S.; Rosca, M.; Figurnov, M.; and Mnih, A. 2020.
\newblock Monte Carlo Gradient Estimation in Machine Learning.
\newblock \emph{Journal of Machine Learning Research}, 21(132): 1--62.

\bibitem[{Morais et~al.(2021)Morais, Kuo, Thomas, T{\"u}ske, and
  Kingsbury}]{morais2021end}
Morais, E.; Kuo, H.-K.~J.; Thomas, S.; T{\"u}ske, Z.; and Kingsbury, B. 2021.
\newblock End-to-end spoken language understanding using transformer networks
  and self-supervised pre-trained features.
\newblock \emph{IEEE International Conference on Acoustics, Speech and Signal
  Processing (ICASSP)}.

\bibitem[{Ondel, Burget, and Cernock{\'{y}}(2016)}]{ondel_variational_2016}
Ondel, L.; Burget, L.; and Cernock{\'{y}}, J. 2016.
\newblock Variational Inference for Acoustic Unit Discovery.
\newblock \emph{SLTU-2016, 5th Workshop on Spoken Language Technologies for
  Under-resourced languages}.

\bibitem[{Ouali, Hudelot, and Tami(2020)}]{ouali2020overview}
Ouali, Y.; Hudelot, C.; and Tami, M. 2020.
\newblock An overview of deep semi-supervised learning.
\newblock \emph{arXiv:2006.05278}.

\bibitem[{Park et~al.(2019)Park, Chan, Zhang, Chiu, Zoph, Cubuk, and
  Le}]{park2019specaugment}
Park, D.~S.; Chan, W.; Zhang, Y.; Chiu, C.-C.; Zoph, B.; Cubuk, E.~D.; and Le,
  Q.~V. 2019.
\newblock SpecAugment: A Simple Data Augmentation Method for Automatic Speech
  Recognition.
\newblock \emph{INTERSPEECH}.

\bibitem[{Pasad et~al.(2021)Pasad, Wu, Shon, Livescu, and Han}]{pasad2021use}
Pasad, A.; Wu, F.; Shon, S.; Livescu, K.; and Han, K.~J. 2021.
\newblock On the Use of External Data for Spoken Named Entity Recognition.
\newblock \emph{arXiv:2112.07648}.

\bibitem[{Pascual et~al.(2019)Pascual, Ravanelli, Serrà, Bonafonte, and
  Bengio}]{Pascual2019}
Pascual, S.; Ravanelli, M.; Serrà, J.; Bonafonte, A.; and Bengio, Y. 2019.
\newblock Learning Problem-Agnostic Speech Representations from Multiple
  Self-Supervised Tasks.
\newblock \emph{INTERSPEECH}.

\bibitem[{Pathak et~al.(2016)Pathak, Krahenbuhl, Donahue, Darrell, and
  Efros}]{pathak2016context}
Pathak, D.; Krahenbuhl, P.; Donahue, J.; Darrell, T.; and Efros, A.~A. 2016.
\newblock Context encoders: Feature learning by inpainting.
\newblock \emph{IEEE Conference on Computer Vision and Pattern Recognition
  (CVPR)}.

\bibitem[{Ravanelli et~al.(2020)Ravanelli, Zhong, Pascual, Swietojanski,
  Monteiro, Trmal, and Bengio}]{Ravanelli2020}
Ravanelli, M.; Zhong, J.; Pascual, S.; Swietojanski, P.; Monteiro, J.; Trmal,
  J.; and Bengio, Y. 2020.
\newblock {Multi-task self-supervised learning for Robust Speech Recognition}.
\newblock \emph{IEEE International Conference on Acoustics, Speech and Signal
  Processing (ICASSP)}.

\bibitem[{Renshaw et~al.(2015)Renshaw, Kamper, Jansen, and
  Goldwater}]{renshaw2015comparison}
Renshaw, D.; Kamper, H.; Jansen, A.; and Goldwater, S. 2015.
\newblock A comparison of neural network methods for unsupervised
  representation learning on the {Zero Resource Speech Challenge}.
\newblock \emph{INTERSPEECH}.

\bibitem[{Rezende, Mohamed, and Wierstra(2014)}]{rezende_stochastic_2014}
Rezende, D.~J.; Mohamed, S.; and Wierstra, D. 2014.
\newblock Stochastic {Backpropagation} and {Approximate} {Inference} in {Deep}
  {Generative} {Models}.
\newblock \emph{International Conference on Machine Learning (ICML)}.

\bibitem[{Riviere et~al.(2020)Riviere, Joulin, Mazar{\'e}, and
  Dupoux}]{riviere2020unsupervised}
Riviere, M.; Joulin, A.; Mazar{\'e}, P.-E.; and Dupoux, E. 2020.
\newblock Unsupervised pretraining transfers well across languages.
\newblock \emph{IEEE International Conference on Acoustics, Speech and Signal
  Processing (ICASSP)}.

\bibitem[{Schatz et~al.(2013)Schatz, Peddinti, Bach, Jansen, Hermansky, and
  Dupoux}]{schatz2013evaluating}
Schatz, T.; Peddinti, V.; Bach, F.; Jansen, A.; Hermansky, H.; and Dupoux, E.
  2013.
\newblock Evaluating speech features with the minimal-pair ABX task: Analysis
  of the classical MFC/PLP pipeline.
\newblock \emph{INTERSPEECH}.

\bibitem[{Schatz et~al.(2014)Schatz, Peddinti, Cao, Bach, Hermansky, and
  Dupoux}]{schatz2014evaluating}
Schatz, T.; Peddinti, V.; Cao, X.-N.; Bach, F.; Hermansky, H.; and Dupoux, E.
  2014.
\newblock Evaluating speech features with the Minimal-Pair ABX task (II):
  Resistance to noise.
\newblock \emph{INTERSPEECH}.

\bibitem[{Schneider et~al.(2019)Schneider, Baevski, Collobert, and
  Auli}]{schneider2019wav2vec}
Schneider, S.; Baevski, A.; Collobert, R.; and Auli, M. 2019.
\newblock wav2vec: Unsupervised Pre-Training for Speech Recognition.
\newblock \emph{INTERSPEECH}.

\bibitem[{Shon et~al.(2021)Shon, Pasad, Wu, Brusco, Artzi, Livescu, and
  Han}]{shon2021slue}
Shon, S.; Pasad, A.; Wu, F.; Brusco, P.; Artzi, Y.; Livescu, K.; and Han, K.~J.
  2021.
\newblock SLUE: New Benchmark Tasks for Spoken Language Understanding
  Evaluation on Natural Speech.
\newblock \emph{arXiv:2111.10367}.

\bibitem[{Simonyan and Zisserman(2015)}]{simonyan2014very}
Simonyan, K.; and Zisserman, A. 2015.
\newblock Very deep convolutional networks for large-scale image recognition.
\newblock \emph{International Conference on Learning Representations (ICLR)}.

\bibitem[{Smolensky(1986)}]{smolensky_parallel_1987}
Smolensky, P. 1986.
\newblock Chapter 6: Information Processing in Dynamical Systems: Foundations
  of Harmony Theory.
\newblock In D.E.~Rumelhart, J.~M., ed., \emph{Parallel Distributed Processing:
  Explorations in the Microstructure of Cognition: Foundations}, chapter~6,
  194--281. MIT Press.

\bibitem[{Soong, Rosenberg, and Juang(1985)}]{soong_vector_1985}
Soong, F.; Rosenberg, A.; and Juang, L. R.~B. 1985.
\newblock A vector Quantization approach to Speaker Recognition.
\newblock \emph{IEEE International Conference on Acoustics, Speech and Signal
  Processing (ICASSP)}.

\bibitem[{Szegedy et~al.(2015)Szegedy, Liu, Jia, Sermanet, Reed, Anguelov,
  Erhan, Vanhoucke, and Rabinovich}]{szegedy2015going}
Szegedy, C.; Liu, W.; Jia, Y.; Sermanet, P.; Reed, S.; Anguelov, D.; Erhan, D.;
  Vanhoucke, V.; and Rabinovich, A. 2015.
\newblock Going deeper with convolutions.
\newblock \emph{IEEE Conference on Computer Vision and Pattern Recognition
  (CVPR)}.

\bibitem[{Sønderby et~al.(2016)Sønderby, Raiko, Maaløe, Sønderby, and
  Winther}]{sonderby_ladder_2016}
Sønderby, C.~K.; Raiko, T.; Maaløe, L.; Sønderby, S.~K.; and Winther, O.
  2016.
\newblock Ladder {Variational} {Autoencoders}.
\newblock \emph{Neural Information Processing System (NeurIPS)}.

\bibitem[{Tagliasacchi et~al.(2020)Tagliasacchi, Gfeller, de~Chaumont~Quitry,
  and Roblek}]{tagliasacchi2020pre}
Tagliasacchi, M.; Gfeller, B.; de~Chaumont~Quitry, F.; and Roblek, D. 2020.
\newblock Pre-training audio representations with self-supervision.
\newblock \emph{IEEE Signal Processing Letters}, 27: 600--604.

\bibitem[{Talnikar et~al.(2021)Talnikar, Likhomanenko, Collobert, and
  Synnaeve}]{talnikar2021joint}
Talnikar, C.; Likhomanenko, T.; Collobert, R.; and Synnaeve, G. 2021.
\newblock Joint masked cpc and ctc training for asr.
\newblock \emph{IEEE International Conference on Acoustics, Speech and Signal
  Processing (ICASSP)}.

\bibitem[{Tsai et~al.(2020)Tsai, Wu, Salakhutdinov, and Morency}]{tsai2020self}
Tsai, Y.-H.~H.; Wu, Y.; Salakhutdinov, R.; and Morency, L.-P. 2020.
\newblock Self-supervised Learning from a Multi-view Perspective.
\newblock \emph{International Conference on Learning Representations (ICLR)}.

\bibitem[{van~den Oord et~al.(2016)van~den Oord, Dieleman, Zen, Simonyan,
  Vinyals, Graves, Kalchbrenner, Senior, and Kavukcuoglu}]{oord_wavenet:_2016}
van~den Oord, A.; Dieleman, S.; Zen, H.; Simonyan, K.; Vinyals, O.; Graves, A.;
  Kalchbrenner, N.; Senior, A.; and Kavukcuoglu, K. 2016.
\newblock {WaveNet}: {A} {Generative} {Model} for {Raw} {Audio}.
\newblock \emph{{Proceedings} of the 9th {ISCA} {Speech} {Synthesis}
  {Workshop}}.

\bibitem[{van~den Oord, Li, and Vinyals(2018)}]{oord2018representation}
van~den Oord, A.; Li, Y.; and Vinyals, O. 2018.
\newblock Representation learning with contrastive predictive coding.
\newblock \emph{arXiv:1807.03748}.

\bibitem[{van~den Oord, Vinyals, and Kavukcuoglu(2018)}]{oord_neural_2018}
van~den Oord, A.; Vinyals, O.; and Kavukcuoglu, K. 2018.
\newblock Neural {Discrete} {Representation} {Learning}.
\newblock \emph{Neural Information Processing System (NeurIPS)}.

\bibitem[{van Niekerk, Nortje, and Kamper(2020)}]{niekerk_vector_2020}
van Niekerk, B.; Nortje, L.; and Kamper, H. 2020.
\newblock Vector-Quantized Neural Networks for Acoustic Unit Discovery in the
  {ZeroSpeech} 2020 Challenge.
\newblock \emph{INTERSPEECH}.

\bibitem[{Vaswani et~al.(2017)Vaswani, Shazeer, Parmar, Uszkoreit, Jones,
  Gomez, Kaiser, and Polosukhin}]{vaswani2017attention}
Vaswani, A.; Shazeer, N.; Parmar, N.; Uszkoreit, J.; Jones, L.; Gomez, A.~N.;
  Kaiser, {\L}.; and Polosukhin, I. 2017.
\newblock Attention is all you need.
\newblock \emph{Neural Information Processing System (NeurIPS)}.

\bibitem[{Versteegh et~al.(2015)Versteegh, Thiolliere, Schatz, Cao, Anguera,
  Jansen, and Dupoux}]{versteegh2015zero}
Versteegh, M.; Thiolliere, R.; Schatz, T.; Cao, X.~N.; Anguera, X.; Jansen, A.;
  and Dupoux, E. 2015.
\newblock The {Zero Resource Speech Challenge} 2015.
\newblock \emph{INTERSPEECH}.

\bibitem[{Wang et~al.(2021)Wang, Wu, Qian, Kumatani, Liu, Wei, Zeng, and
  Huang}]{wang2021unispeech}
Wang, C.; Wu, Y.; Qian, Y.; Kumatani, K.; Liu, S.; Wei, F.; Zeng, M.; and
  Huang, X. 2021.
\newblock Unispeech: Unified speech representation learning with labeled and
  unlabeled data.
\newblock \emph{International Conference on Machine Learning (ICML)}.

\bibitem[{Wang, Tang, and Livescu(2020)}]{wang2020unsupervised}
Wang, W.; Tang, Q.; and Livescu, K. 2020.
\newblock Unsupervised pre-training of bidirectional speech encoders via masked
  reconstruction.
\newblock \emph{IEEE International Conference on Acoustics, Speech and Signal
  Processing (ICASSP)}.

\bibitem[{Wang, Chung, and Lee(2017)}]{wang2017gate}
Wang, Y.-H.; Chung, C.-T.; and Lee, H.-y. 2017.
\newblock Gate activation signal analysis for gated recurrent neural networks
  and its correlation with phoneme boundaries.
\newblock \emph{INTERSPEECH}.

\bibitem[{Wiskott and Sejnowski(2002)}]{wiskott2002slow}
Wiskott, L.; and Sejnowski, T.~J. 2002.
\newblock Slow feature analysis: Unsupervised learning of invariances.
\newblock \emph{Neural Computation}, 14(4): 715--770.

\bibitem[{Yang et~al.(2021)Yang, Chi, Chuang, Lai, Lakhotia, Lin, Liu, Shi,
  Chang, Lin et~al.}]{yang2021superb}
Yang, S.-w.; Chi, P.-H.; Chuang, Y.-S.; Lai, C.-I.~J.; Lakhotia, K.; Lin,
  Y.~Y.; Liu, A.~T.; Shi, J.; Chang, X.; Lin, G.-T.; et~al. 2021.
\newblock SUPERB: Speech processing Universal PERformance Benchmark.
\newblock \emph{INTERSPEECH}.

\end{thebibliography}

\end{document}